\documentclass[11pt,a4paper]{article}
\usepackage{jheppub}
    \makeatletter
\def\@fpheader{\relax}
\makeatother
\usepackage{mathtools}
\usepackage{graphicx}
	\graphicspath{ {images/} }
\usepackage{blindtext}	
\usepackage{amsmath,amssymb,amsthm}
\usepackage{latexsym,graphicx,color}%,subfigure}
\usepackage{enumerate}
\usepackage{color,ulem}

\usepackage{hyperref}
\newcommand{\non}{\nonumber\\}
\newcommand{\D}{\mathcal{D}}

\newcommand{\Tr}{{\rm Tr}}
\newcommand{\diag}{{\rm diag}}

\newcommand{\bea}{\begin{eqnarray}}
\newcommand{\eea}{\end{eqnarray}}
\def\Tr{ \hbox{\rm Tr}}

\def\changed{\textcolor{black}}

\def\6#1{{\underline{#1}}}
\def\m6#1{{\underline{#1}\,}}

\newdimen\Tdim
\def\ispan{{\setbox0=\hbox{i}%
\Tdim\ht0\advance\Tdim\dp0\rule[-\dp0]{0pt}{\Tdim}}}
\def\jspan{{\setbox0=\hbox{j}%
\Tdim\ht0\advance\Tdim\dp0\rule[-\dp0]{0pt}{\Tdim}}}
\def\Tspan#1{{\setbox0=\hbox{#1}%
\Tdim\ht0\advance\Tdim\dp0\advance\Tdim.55ex\rule[-\dp0]{0pt}{\Tdim}\box0}}

\def\be{\begin{eqnarray}}
\def\ben{\begin{eqnarray*}}
\def\ee{\end{eqnarray}}
\def\een{\end{eqnarray*}}
\def\Tr{{\rm Tr}}

\def\D{\mathcal{D}}

\def\=:{=\hspace{-.7em}\raisebox{1.1ex}{.}\hspace{.1em}\raisebox{-0.2ex}{.} }

\newcommand {\1}[1]{\frac{1}{#1}}

\newcommand {\beq}{\begin{eqnarray}}
\newcommand {\eeq}{\end{eqnarray}}

\newcommand{\dd}{\langle dd \rangle}

\preprint{YGHP-21-01}
\title{{\bf Chiral non-Abelian vortices 
and their confinement in three flavor dense QCD
}}

\author[a,b]{Minoru~Eto,}
\author[b,c]{and Muneto~Nitta} 

\affiliation[a]{Department of Physics, Yamagata University, Kojirakawa-machi 1-4-12, Yamagata, Yamagata 990-8560, Japan}
\affiliation[b]{Research and Education Center for Natural Sciences, Keio University, 4-1-1 Hiyoshi, Yokohama, Kanagawa 223-8521, Japan}
\affiliation[c]{Department of Physics, Keio University, 4-1-1 Hiyoshi, Kanagawa 223-8521, Japan}

\emailAdd{meto(at)sci.kj.yamagata-u.ac.jp}
\emailAdd{nitta(at)phys-h.keio.ac.jp}

\date{\today}

\abstract{
We find chiral non-Abelian vortices 
\changed{
having 
windings only in one of 
the diquark condensations 
of left-handed and right-handed quarks}  
in the color-flavor locked phase of dense QCD. 
They are the minimum vortices 
carrying half color magnetic fluxes 
of those of non-Abelian semi-superfluid vortices 
(color magnetic flux tubes) 
and 1/6 quantized superfluid circulations 
of Abelian superfluid vortices. 
These vortices carry 
${\mathbb C}P^2$ orientational moduli in the internal space 
corresponding to their fluxes. 
The ${\mathbb C}P^2$ moduli of 
two chiral non-Abelian vortices with 
chiralities opposite to each other 
are energetically favored to be aligned 
while those of a vortex and anti-vortex to be orthogonal,
and then these vortices attract each other.
They are attached by chiral domain walls 
in the presence of the mass and axial anomaly terms explicitly breaking 
axial and chiral symmetries.
\changed{We numerically show that} two chiral non-Abelian vortices with chiralities opposite to each other  
are connected by a chiral domain wall, consisting 
a mesonic bound state
which is nothing but 
a non-Abelian semi-superfluid vortex. 
We also show that 
Abelian and non-Abelian axial vortices 
attached by chiral domain walls are all unstable to decay into 
a set of chiral non-Abelian vortices.
Furthermore, we find that 
chiral non-Abelian vortices exhibit unique features:
one is the so-called 
topological obstruction implying that 
unbroken symmetry generators in the bulk are not defined globally around the vortices,
and the other is 
  color non-singlet Aharonov-Bohm (AB) phases 
implying that 
quarks encircling these vortices can detect the colors of magnetic fluxes of them at infinite distances.  
}

\begin{document}

\maketitle

\section{Introduction}
What are states of matter at extreme conditions is
one of the challenging problem in modern physics. 
The ground state of the cold QCD matter at
high densities is expected to exhibit color superconductivity, 
which may be realized in cores of neutron stars~\cite{Alford:2007xm}.  
Various phases have been proposed  for color superconductivity; 
 the color-flavor locked (CFL) phase~\cite{Alford:1998mk} 
 in three-flavor symmetric matter 
is realized extremely high density limit, 
while 
the two-flavor superconducting (2SC) phase~\cite{Alford:1997zt, Rapp:1997zu}
was also proposed for two-flavor symmetric matter.  
If a color superconductor is realized in the core of neutron stars,
there must appear quantum vortices, 
{\it i.~e.~} vortices with quantized circulations, 
because of rapid rotations.
In color-superconducting quark matter,  
quantum vortices or color magnetic flux tubes appear,  
as reviewed in Ref.~\cite{Eto:2013hoa}. 
In the CFL phase, Abelian superfluid vortices 
are created by rotations \cite{Forbes:2001gj,Iida:2002ev},  
which are dynamically unstable to decay into 
more stable vortices  
\cite{Nakano:2007dr,Cipriani:2012hr,Alford:2016dco}.
The most stable vortices are 
non-Abelian semi-superfluid vortices carrying color magnetic fluxes and
1/3 circulation of the Abelian superfluid
vortices~\cite{Balachandran:2005ev, Nakano:2007dr, Nakano:2008dc,
  Eto:2009kg, Eto:2013hoa}, 
  \changed{
  which are analogous to non-Abelian vortices in 
  supersymmetric QCD \cite{Hanany:2003hp,Auzzi:2003fs,
Hanany:2004ea,Shifman:2004dr,Eto:2004rz,Eto:2005yh} 
(see Refs.~\cite{Tong:2005un,Eto:2006pg,Shifman:2007ce,Shifman:2009zz} as a review) 
and two-Higgs doublet models \cite{Eto:2018hhg,Eto:2018tnk,
Eto:2019hhf,Eto:2020hjb,Eto:2020opf}.
}
  A non-Abelian vortex confines massless particles in its core; 
  One is bosonic 
   Nambu-Goldstone 
   ${\mathbb C}P^2$ modes originated from 
  spontaneous breaking of the CFL symmetry 
   in its core 
   \cite{Nakano:2007dr,Eto:2009bh,Eto:2013hoa,Eto:2009tr},
   and the other is
  gapless Majorana fermions with more topological origin 
  \cite{Yasui:2010yw,Fujiwara:2011za,Chatterjee:2016ykq}. 
  \changed{
  Under a rapid rotation, there appear a huge number 
  of vortices 
  (about $10^{19}$ for typical neutron stars). 
  They will form a vortex lattice 
  \cite{Kobayashi:2013axa}  
  that behaves as a polarizer of photons 
  \cite{Hirono:2012ki}.
  }
One of the most recent progress is 
vortices penetrating through crossover between the CFL phase and hyperon nuclear matter 
within a quark-hadron continuity~\cite{Alford:2018mqj, Chatterjee:2018nxe,
  Chatterjee:2019tbz, Cipriani:2012hr, Cherman:2018jir, Hirono:2018fjr, Hirono:2019oup, Cherman:2020hbe}. 
  \changed{
While it was suggested that 
one superfluid vortex in the hyperon nuclear matter 
is connected to one non-Abelian vortex 
in the CFL phase \cite{Alford:2018mqj}, 
it was proved in Refs.~\cite{Chatterjee:2018nxe,Chatterjee:2019tbz}
that three superfluid vortices 
meet three non-Abelian vortices 
at a point called a Boojum
 \cite{Cipriani:2012hr}.
}

\changed{
The CFL phase is characterized by the two diquark condensations 
of left and right-handed quarks $q_{\rm L,R}$,
$(\Phi_{\rm L,R})_{\alpha a} \sim \epsilon_{\alpha \beta \gamma} 
\epsilon_{abc} q_{\rm L,R}^{\beta b}  q_{\rm L,R}^{\gamma c}$.  
with the color indices $\alpha,\beta,\gamma=r,g,b$ 
and the flavor indices $a,b,c=u,d,s$.   
In the ground states, they
both develop VEVs 
as $\Phi_{\rm L} = - \Phi_{\rm R}$. 
Thus, with defining $\Phi \equiv \Phi_{\rm L} = - \Phi_{\rm R}$,  
one has discussed non-Abelian vortices 
in terms of $\Phi$ and gauge fields. 
However, since the relation $\Phi_{\rm L} = - \Phi_{\rm R}$ 
holds only in the ground state, we do not have to assume it 
for excited states such as vortices.  
In fact, 
a similar situation can be found in 
two-component condensed matter systems: 
two-gap superconductors 
 \cite{Babaev:2001hv,
 doi:10.1143/JPSJ.70.2844,
PhysRevLett.88.017002,Goryo_2007} 
and 
two-component Bose-Einstein condensates (BECs)
\cite{
Son:2001td,
Kasamatsu:2004tvg,
Cipriani:2013nya,
Tylutki:2016mgy,
Eto:2017rfr,
Eto:2019uhe,Kobayashi:2018ezm}.
In these systems, there are two condensations 
$\Phi_1$ and $\Phi_2$.
For a singly quantized vortex 
both fields have the winding $\Phi_1= \Phi_2 \sim e^{i\varphi}$ 
with azimuthal angle $\varphi$. 
However, they also admit so-called half-quantum vortices 
 $(\Phi_1,\Phi_2 )\sim (e^{i\varphi},1)$ 
 or   $(\Phi_1,\Phi_2 )\sim (1, e^{i\varphi})$,
 denoted as $(1,0)$ or $(0,1)$, respectively.
 It is called half-quantum since it carries 
 a half magnetic flux in superconductors 
 or half circulation in BECs.\footnote{
 \changed{
 Strictly speaking, they are half quantized when the VEVs of 
 $\Phi_1$ and $\Phi_2$ are the same. 
 When their VEVs $\left< \Phi_1 \right> = v_1$, 
 $\left< \Phi_2 \right> =v_2$ are different, they are fractionally quantized as $v_1^2/(v_1^2 + v_2^2)$ and 
 $v_2^2/(v_1^2 + v_2^2)$.
 } 
 }
When the system contains an 
interaction term $\Phi_1^* \Phi_2 + {\rm c.c.}$ 
known as a Josephson term in superconductors 
or Rabi coupling in BECs,
$(1,0)$ and $(0,1)$ vortices are connected by 
a sine-Gordon soliton 
 \cite{doi:10.1143/JPSJ.70.2844,
PhysRevLett.88.017002,Son:2001td}\footnote{
\changed{
While the Josephson coupling is inevitable in superconductors, 
the Rabi coupling is in general absent in BECs and 
one can introduce it  as 
an experimentally controllable parameter.
}};
they are confined 
to form a singly quantized vortex. 
When there are no such terms, they are deconfined 
and are weakly interacting each other.
Thus,
one may wonder if the same can be considered 
for vortices in the CFL phase.
}

\changed{
In this paper, we investigate non-Abelian vortices having windings 
only in left $\Phi_{\rm L}$ or right $\Phi_{\rm R}$ condensation, 
while the previously known 
non-Abelian vortices 
have windings in the both components simultaneously. 
We call them ``chiral non-Abelian vortices'' 
in the sense that quarks of only left or right chirality 
participate in the vortices.  
A single non-Abelian semi-superfluid vortex 
can be decomposed into 
chiral non-Abelian vortices of chiralities opposite to each other.
Chiral non-Abelian vortices
carry half color magnetic fluxes 
of those of non-Abelian semi-superfluid vortices 
and 1/6 quantized superfluid circulations 
of Abelian superfluid vortices.  
%Each of which carries a half color magnetic flux of that of one non-Abelian semi-superfluid vortex, 
%that is the same amount with that of 
%a non-Abelian Alice string in the 2SC + $\dd$ phase.
We find that a single chiral non-Abelian vortex 
carries ${\mathbb C}P^2$ orientational moduli in the internal space  corresponding to 
its color magnetic flux. 
As the case of the Josephson or Rabi term for two-component condensed matter systems, 
a chiral non-Abelian vortex is attached by 
a chiral domain wall \cite{Eto:2013hoa,Eto:2013bxa}   
in the presence of mass and axial anomaly terms explicitly breaking 
axial and chiral symmetries. 
We also 
study energetics of two chiral vortices 
in the absence of mass and axial anomaly terms.
In the coexistence of a set of two chiral vortices 
with opposite chiralities, 
 their ${\mathbb C}P^2$ moduli must be aligned energetically 
 and they attract each other. 
 On the other hand, 
 in the  coexistence of a chiral vortex and an anti-chiral vortex, 
their ${\mathbb C}P^2$ moduli are  energetically orthogonal to each other,   and they attract each other.
% Then, the two chiral non-Abelian vortices attract each other since  the tension of each chiral non-Abelian vortex  coincides with that of a single non-Abelian semi-superfluid vortex. 
We show that in the presence of mass or axial anomaly term,
chiral non-Abelian vortices of opposite chirality are connected  
by a chiral domain wall 
and are linearly confined. 
}

\changed{
Other interesting features that we find in this paper 
are so-called {\it topological obstruction} 
(see Appendix \ref{sec:topological-obstruction})  
\cite{Schwarz:1982ec,
  Alford:1990mk, Alford:1990ur, Alford:1992yx, Preskill:1990bm, 
  Bucher:1992bd, Lo:1993hp,  
Bolognesi:2015mpa} 
and non-Abelian Aharanov-Bohm (AB) phases.
First,
the topological obstruction 
that chiral non-Abelian vortices exhibit  
implies that generators of the unbroken symmetry in the ground state 
are not globally defined around the vortices. 
Second, 
}the chiral non-Abelian vortices 
exhibit color non-singlet (generalized) AB phases 
so that the quarks can detect the colors of magnetic fluxes of these vortices 
at large distances. 
%In the confined phase, 
The bound state of two chiral non-Abelian vortices with the opposite chiralities, 
\changed{
equivalent to a single non-Abelian semi-superfluid vortex 
at large distance,  
}
exhibits only color singlet (generalized) AB phases 
so that the quarks cannot detect the color magnetic flux of such a bound state  
at large distances.

\changed{
Finally, we will point out that 
chiral non-Abelian vortices are 
the most fundamental elements among 
topological solitons 
formed during the chiral symmetry breaking.
The CFL
phase is accompanied by spontaneous breaking of the chiral symmetry since the diquark condensations 
$\Phi_{\rm L,R}$
%_{\alpha a} \sim \epsilon_{\alpha \beta \gamma} \epsilon_{abc} q_{\rm L,R}^{\beta b}  q_{\rm L,R}^{\gamma c}$  
of left and right-handed quarks $q_{\rm L,R}$
both develop VEVs. 
This breaking admits several vortices 
without color magnetic fluxes.
}
%, with the color indices $\alpha,\beta,\gamma=r,g,b$ and the flavor indices $a,b,c=u,d,s$.   
%Chiral non-Abelian vortices that we find in this paper are related with solitons in the chiral symmetry breaking as follows. 
When the $U(1)_{\rm A}$ axial symmetry is spontaneously broken, 
it admits an axial vortex winding around $U(1)_{\rm A}$.
This vortex is attached by $2N$ domain walls because of 
the anomaly term explicitly breaking 
the $U(1)_{\rm A}$ axial symmetry, 
in contrast to an analogous axial string in
the linear sigma model for chiral symmetry breaking  
which is  
attached by $N$ domain walls 
\cite{Balachandran:2001qn}.
The $U(N)_{\rm L}\times U(N)_{\rm R}$ chiral symmetry breaking admits 
a non-Abelian axial string, 
which is  
attached by one (or two) chiral domain wall(s) 
depending on the form of mass terms, 
analogous to one in the linear sigma model for chiral symmetry breaking  
\cite{Balachandran:2002je,Nitta:2007dp,Nakano:2007dq,Nakano:2008dc,Eto:2009wu}.
The $U(1)_{\rm A}$ Abelian axial vortex mentioned above 
is dynamically split into $N$ non-Abelian axial strings 
\changed{
by domain wall tensions, 
where 
each non-Abelian string 
}
is attached by one (or two) chiral domain wall(s),
as an analogous decay was studied 
in the linear sigma model for chiral symmetry breaking 
\cite{Eto:2013hoa,Eto:2013bxa}.
\changed{
We find that a single non-Abelian axial string decays 
into a pair of a chiral non-Abelian vortex and an anti-chiral non-Abelian vortex, 
while a single Abelian axial string decays into a set of
$N$ chiral non-Abelian vortices and $N$ anti-chiral non-Abelian vortices. 
Thus, chiral non-Abelian vortices are the most fundamental strings.
}\footnote{
The chiral non-Abelian vortex can be considered as a hybrid of 
a non-Abelian semi-superfluid vortex (color flux tube) 
and a non-Abelian axial string:
the former winds around $\Phi_{\rm L}$ and $\Phi_{\rm R}$ 
with the same windings,
the latter winds around them with the opposite windings,
and 
the chiral non-Abelian vortex winds around 
only either of  $\Phi_{\rm L}$ and $\Phi_{\rm R}$, achieved by 
 a half non-Abelian semi-superfluid vortex (color flux tube) 
and a half non-Abelian axial string.
}

%**************

This paper is organized as follows.
In Sec.~\ref{sec:GL}, we review the Ginzburg-Landau (GL) theory 
paying attention to symmetries, 
and give the order parameter manifold (OPM) 
which is a new result.
In Sec.~\ref{sec:superfluid-vrtx}, we review superfluid vortices: 
Abelian $U(1)_{\rm B}$ superfluid vortices 
and non-Abelian semi-superfluid vortices (color flux tubes). 
In Sec.~\ref{sec:axial-chiral-vtx}, we discuss 
Abelian and non-Abelian axial vortices. 
In Sec.~\ref{sec:chiralNA}, we construct chiral non-Abelian vortices 
in the absence of chiral symmetry breaking terms, 
and show that they exhibit the topological obstruction and (generalized) AB phases of 
quarks encircling them.
In Sec.~\ref{sec:energy}, 
we discuss energetics of 
a single chiral non-Abelian vortex, 
non-Abelian semi-superfluid vortex, 
non-Abelian axial vortex, 
and more general composite vortices.
We find that the 
${\mathbb C}P^2$ orientations 
of two chiral vortices with the opposite chiralities  
are energetically favored 
to be aligned to each other
and then they attract each other, 
while those of chiral vortex and anti-vortex 
 with the opposite chiralities  
 are energetically favored to be orthogonal to each other 
and then they attract each other. 
In Sec.~\ref{sec:chiral-sym-br}, 
we show that Abelian and non-Abelian axial 
vortices are attached by chiral domain walls 
in the presence of axial and chiral symmetry breaking terms, 
and discuss decay of these vortices.
In Sec.~\ref{sec:chiralNAmolecule}, we construct 
a mesonic bound state of two chiral non-Abelian vortices with the opposite chiralities.
Sec.~\ref{sec:summary} is devoted to a summary and discussion.
\changed{
In Appendix \ref{sec:term}, the terminologies used in this paper are summarized.
}
In Appendix \ref{sec:OPM}, we give detailed discussions 
on symmetry breakings in the CFL phase, 
and determine associated OPMs.
\changed{
In Appendix \ref{sec:two-flavors}, 
chiral non-Abelian vortices in the CFL phase 
are compared with non-Abelian Alice strings 
\cite{Fujimoto:2020dsa,Fujimoto:2021bes,Fujimoto:2021wsr} 
in the 2SC + $dd$ phase of two-flavor quark matter proposed recently \cite{Fujimoto:2019sxg,Fujimoto:2020cho}.
}

%%%%%%%%%%%%%
\section{Color-flavor locked phase of three flavor quark matter}\label{sec:GL}

In this section, after we review the color-flavor locked phase of dense QCD, we give OPMs and their topology as a new result.

The (approximate) symmetry of $N$ flavor quark matter is
(up to discrete groups)  
\beq
G =
SU(N)_{\rm C} \times
U(1)_{\rm B} \times U(1)_{\rm A} \times 
SU(N)_{\rm L} \times SU(N)_{\rm R}
\eeq
where 
$SU(N)_{\rm C}$ is the color gauge group, 
and the rests are global symmetries: 
$U(1)_{\rm B}$, 
$U(1)_{\rm A}$
and $SU(N)_{\rm L} \times SU(N)_{\rm R}$ are 
baryon number, axial, 
and chiral symmetries, respectively. 
See Appendix \ref{sec:OPM} for more precise description 
including discrete groups.  
The light quarks 
$q_{\rm L,R} = (q_{\rm L,R})_{\alpha a}$ with 
$\alpha = 1,2,\cdots,N$ ($\alpha = r,g,b$ for $N=3$), 
$a = 1,2,\cdots,N$ ($a= u,d,s$ for $N=3$)
and heavy quarks 
$Q_{\rm L,R}= (Q_{\rm L,R})_{\alpha}$  transform under $G$ as 
\beq
&& q_{\rm L} \to e^{i\theta_{\rm B}/2} e^{i\theta_{\rm A}/2} g_{\rm C}^* \, q_{\rm L}\, U_{\rm L}^T ,
\quad
 q_{\rm R} \to e^{i\theta_{\rm B}/2} e^{-i\theta_{\rm A}/2} g_{\rm C}^* \, q_{\rm R}\, U_{\rm R}^T \non
&& Q_{\rm L} \to 
g_{\rm C}^* \,q_{\rm L} ,\quad
 Q_{\rm R} \to 
g_{\rm C}^* \,q_{\rm R}  ,
\eeq
where we have not introduced heavy quark flavor symmetry, 
and have assigned no $U(1)_{\rm B}$ and $U(1)_{\rm A}$ charges on the heavy quarks.\footnote{The anti-fundamental representation 
$*$ of quarks for the color group is a convention to make the representation of the condensations 
$\Phi_{\rm L,R}$ introduced below to be fundamental 
in Eq.~(\ref{eq:G-on-Phi}), below.
The situation in our mind is that the only light quarks are condensed by forming diquark pairs
while the heavy quarks are not.
Thus, we use the terminology ``$U(1)_{\rm B}$ baryon symmetry'' 
for the spontaneously broken 
baryon symmetry associated only with the light quarks 
with no charges for the heavy quarks,
while the heavy quarks are also charged 
under the conventional baron symmetry.
}
The case of $N=3$ corresponds to the CFL phase 
of dense QCD, in which case the light quarks 
constitute diquark condensations as
$(\Phi_{\rm L,R})_{\alpha a} \sim \epsilon_{\alpha \beta \gamma} 
\epsilon_{abc} q_{\rm L,R}^{\beta b}  q_{\rm L,R}^{\gamma c}$. 
Hereafter, we 
mostly consider the condensates
$(\Phi_{\rm L,R})_{\alpha a}$ as $N$ by $N$ matrices of complex scalar fields
on which the symmetries $G$ act as
\beq
 && \Phi_{\rm L} \to 
 e^{i\theta_{\rm B} +i\theta_{\rm A}} g_{\rm C} \Phi_{\rm L} U_{\rm L}^\dagger , \quad
  \Phi_{\rm R} \to 
   e^{i\theta_{\rm B} - i \theta_{\rm A}} 
  g_{\rm C} \Phi_{\rm R} U_{\rm R}^\dagger \nonumber\\
  && g_{\rm C} \in SU(N)_{\rm C}, \quad 
  U_{\rm L,R} \in SU(N)_{\rm L,R}, \quad
   e^{i\theta_{\rm B}}\in U(1)_{\rm B} ,\quad
    e^{i\theta_{\rm A}} \in U(1)_{\rm A}. \label{eq:G-on-Phi}
\eeq
The vector symmetry $SU(N)_{\rm L+R}$ given by $U_{\rm L} = U_{\rm R}$
is a subgroup of the chiral symmetry 
$SU(N)_{\rm L} \times SU(N)_{\rm R}$, and
the rest of generators outside $SU(N)_{\rm L+R}$ defines the coset space
$[SU(N)_{\rm L} \times SU(N)_{\rm R}] / SU(N)_{\rm L+R}$ 
$\simeq SU(N)$ which we sometimes denote 
$SU(N)_{\rm L-R}$ although this does not form a group.

In this paper, we use the static GL free energy for studying vortices.
The GL Lagrangian for the CFL phase was obtained as
\cite{Giannakis:2001wz,Iida:2000ha,Iida:2001pg}
\beq
{\cal L} &=& \Tr\left[ -\frac{1}{4}F^{ij}F_{ij} + \D_i \Phi_{\rm L}^\dag \D^i \Phi_{\rm L} + \D_i \Phi_{\rm R}^\dag \D^i \Phi_{\rm R} \right] - V,\\
V &=& - \frac{m^2}{2}\Tr[ \Phi_{\rm L}^\dag \Phi_{\rm L} + \Phi_{\rm R}^\dag \Phi_{\rm R}] + \frac{\lambda_1}{4}\Tr[(\Phi_{\rm L}^\dag \Phi_{\rm L} )^2 + (\Phi_{\rm R}^\dag \Phi_{\rm R})^2] \non
&&+ \frac{\lambda_2}{4}( \Tr[\Phi_{\rm L}^\dag \Phi_{\rm L}]^2 + \Tr[\Phi_{\rm R}^\dag \Phi_{\rm R}]^2)
+ \frac{\lambda_3}{2}\Tr[\Phi_{\rm L}^\dag \Phi_{\rm L}] \Tr[\Phi_{\rm R}^\dag \Phi_{\rm R}]
+ 
\frac{\lambda_4}{2}\Tr[ \Phi_{\rm L}\Phi_{\rm L}^\dag 
 \Phi_{\rm R}\Phi_{\rm R}^\dag]
\non
&&+ \left[\gamma_1 \Tr (\Phi_{\rm L}^\dagger \Phi_{\rm R}) 
+ \gamma_2 \Tr [(\Phi_{\rm L}^\dagger \Phi_{\rm R})^2] + 
\gamma_3 \det (\Phi_{\rm L}^\dagger \Phi_{\rm R}) 
+ ({\rm c.c.}) \right],  \label{eq:GL}
\eeq
where the GL coefficients depending on the temperature, density and so on can be found in Refs.~\cite{Giannakis:2001wz,Iida:2000ha,Iida:2001pg}. 
Among the global symmetries, the axial  
$U(1)_{\rm A}$ and chiral symmetries 
are explicitly broken 
in the presence of the last terms 
for $\gamma_{1,2,3} \neq 0$ 
with $U(1)_{\rm B} \times SU(N)_{\rm L+R}$ remaining exaxt.
\changed{
The GL theory is valid only near 
the transition temperature $T_{\rm c}$. 
Beyond the GL theory, we need Bogoliubov-de Gennes (BdG) 
formulation \cite{Yasui:2010yw,Fujiwara:2011za}.
}

The ground state is given by
\beq
\Phi_{\rm L} = -\Phi_{\rm R} = v {\bf 1}_N,\qquad
v \equiv \left(\frac{m^2}{\lambda_1 + N \lambda_2 + N \lambda_3}\right)^{\frac{1}{4}} 
\eeq
for small $\gamma$'s.
The symmetry $G$ is spontaneously broken down to 
the CFL symmetry given by
\beq 
 H = SU(N)_{\rm C+L+R},  \quad
 g_{\rm C} = U_{\rm L} = U_{\rm R}.\label{eq:CFL}
\eeq
The chiral symmetry,
 $U(1)_{\rm A}$ and $U(1)_{\rm B}$ symmetries are spontaneously broken.

According to Appendix \ref{sec:OPM},
the full OPM for the symmetry breaking
can be written, 
with taking into account discrete groups, 
as
\beq
 {\cal M}
 &=& {G\over H} \simeq
 {U(N)_{\rm C-(L+R)+B} \ltimes U(N)_{\rm L-R+A} 
 \over ({\mathbb Z}_2)_{\rm A + B}}
 =
 {{\cal M}_{\rm V} \ltimes {\cal M}_{\rm A}
 \over ({\mathbb Z}_2)_{\rm A + B}}, \label{eq:fullOPM0}
\eeq
where 
$({\mathbb Z}_2)_{\rm A+B}$ is generated by 
$(-1,-1) \in U(1)_{\rm B} \times U(1)_{\rm A}$, 
and $F \ltimes B$ denotes a fiber bundle 
with a fiber $F$ over a base manifold $B$. 
Here, we have defined  
the sub-OPMs for the vector symmetry breaking 
and for the axial and chiral symmetry breakings by
\beq
&& {\cal M}_{\rm V} 
 \simeq U(N)_{\rm C-(L+R)+B} 
 \simeq
 {U(1)_{\rm B}  \times SU(N)_{\rm C-(L+R)} \over 
 ({\mathbb Z}_N)_{\rm C-(L+R)+B} }, \non
&& {\cal M}_{\rm A} 
 \simeq 
 U(N)_{\rm L-R+A} 
 \simeq 
{U(1)_{\rm A} \times  SU(N)_{\rm L-R} \over 
({\mathbb Z}_N)_{\rm L-R+A}  } ,\label{eq:sub-OPM}
\eeq
respectively, with coset spaces
\beq
  SU(N)_{\rm C-(L+R)} \simeq
  {SU(N)_{\rm C} \times SU(N)_{\rm L+R} 
 \over 
SU(N)_{\rm C+L+R} },\quad
SU(N)_{\rm L-R} \simeq
   { SU(N)_{\rm L} \times SU(N)_{\rm R} \over  SU(N)_{\rm L+R}}.
\eeq
For the details of derivation, see Appendix \ref{sec:OPM}.

The $({\mathbb Z}_2)_{\rm A + B}$ in the denominator 
of Eq.~(\ref{eq:fullOPM0})
was not recognized before 
(see Eq.~(2.26) of Ref.~\cite{Eto:2013hoa}),
and 
this is a key point to understand 
chiral non-Abelian vortices found in  this paper.
The nontrivial first homotopy groups of the sub-OPMs
\beq
 \pi_1 ({\cal M}_{\rm V}) \simeq {\mathbb Z}, \quad
 \pi_1 ({\cal M}_{\rm A}) \simeq {\mathbb Z}
\eeq
support non-Abelian semi-superfluid vortices (Sec.~\ref{sec:NAV})
and non-Abelian axial vortices (Sec.~\ref{sec:NAGV}), respectively, 
but they are not the minimum vortices.
On the contrary,
the nontrivial first homotopy group of the full OPM
\beq
 \pi_1 ({\cal M}) \simeq {\mathbb Z}
\eeq
supports chiral non-Abelian vortices 
as the minimum vortices.

Here, one comment is in order.
Considering $N=1$ in the full OPM in Eq.~(\ref{eq:fullOPM0}),
we obtain the OPM for two-component BECs or superconductors, 
see Ref.~\cite{Eto:2011wp}, 
allowing half-quantized vortices.
Thus, our case is a non-Abelian generalization of such two-component
condensed matter systems.

For later conveniences, we define gauge invariants  
\beq
 && \Sigma \equiv \Phi_{\rm R}^\dagger \Phi_{\rm L} 
 \quad \det \Phi_{\rm L}, \quad \det \Phi_{\rm R}  .
 \label{eq:Sigma}
\eeq
Here, $\Sigma$ is the chiral symmetry breaking order parameter. 
These gauge invariants transform under the flavor symmetry as 
\beq
&& \Sigma \to e^{2i\theta_{\rm A}} U_{\rm R} \Sigma U_{\rm L}^\dagger \non
&& \det \Phi_{\rm L} \to 
  e^{N i\theta_{\rm B} + N i \theta_{\rm A}} 
  \det \Phi_{\rm L},  \non
&&  \det \Phi_{\rm R} 
  \to  e^{N i\theta_{\rm B} - N i \theta_{\rm A}} 
  \det \Phi_{\rm R}  .
\eeq

In the following sections, we classify various vortices in the CFL phase 
as summarized in Table \ref{tab:summary}.
To this end, let us introduce labels of vortices by 
\beq 
 (m,n): \quad \det \Phi_{\rm L}\sim e^{i m \varphi}, \quad 
  \det \Phi_{\rm R}\sim e^{i n \varphi}, \label{eq:label}
\eeq
with winding numbers $n$ and $m$ of the gauge invariants 
$\det \Phi_{\rm L}$ and $\det \Phi_{\rm R}$, respectively. 
Here $\varphi$ is the angle coordinate of the polar coordinates 
in two dimensional space perpendicular to the vortex.

%%%   Table   %%%%%
\begin{table*}[t]
  \centering
  \begin{tabular}{c|c|c|c|c|c|c}%||cc|cc|cc|c}
     vortex & OPM & label & 
      \begin{tabular}{c}
      $U(1)_{\rm B}$ \\ circul. \end{tabular} & 
      \begin{tabular}{c}
      color\\
      {\small magnetic} \\
      flux %${\cal F}$
      \end{tabular}
      & \begin{tabular}{c} 
        $U(1)_{\rm A}$ \\{\small winding} 
        \end{tabular}
     & \begin{tabular}{c}  
      chiral \\ circul.\end{tabular}
 \\  \hline \hline
  %%%%%%%%%%%%%
    %%%%%%%%%
    \begin{tabular}{c}
           {pure color magnetic} \\
      flux tube
    \end{tabular}
    & $SU(N)_{\rm C}$ & $(0,0)$ & 0 & 1 & 0 & 0 
    \\ \hline
    %%%%%%%
    \begin{tabular}{c}
    Abelian\\ superfluid 
    vortex \end{tabular}
    & $U(1)_{\rm B}$ & $(N,N)$ & 1 & 0 & 0 & 0 
  \\ \hline
    %%%%%%%%%%%%%%%%
     \begin{tabular}{c}
       NA semi-superfluid \\
      vortex
     \end{tabular}
     & ${\cal M}_{\rm V}$ & $(1,1)$ & $\displaystyle{\frac{1}{N}}$ &  $\displaystyle{\frac{1}{N}}$ & 0 & 0 
\\    \hline
  %%%%%%%%%%%%%
    \begin{tabular}{c}
   Abelian\\
    axial  
    vortex  \end{tabular}
    & $U(1)_{\rm A}$ & $(N,-N)$ & 0 & 0 & 1 & 0 
    \\ \hline
    %%%%%%%%%%%%%%%%
      %%%%%%%%%%%%%
    \begin{tabular}{c}
    NA \\axial
    vortex
    \end{tabular}
    & ${\cal M}_{\rm A}$ & $(1,-1)$ & 0 & 0 &  $\displaystyle{\frac{1}{N}}$ &  $\displaystyle{\frac{1}{N}}$ 
    \\ \hline
    %%%%%%%%%%%%%%%%
    %%%%%%%%%%%%%%%%%%%%%%%%%%%
     \begin{tabular}{c}
      chiral NA \\
      vortex
      \end{tabular}
     & ${\cal M}$ & \begin{tabular}{c} $(1,0)$ \\or\\ $(0,1)$ \end{tabular} 
      &  $\displaystyle{\frac{1}{2N}}$ & $\displaystyle{\frac{1}{2N}}$  & $\displaystyle{\frac{1}{2N}}$  & $\displaystyle{\frac{1}{2N}}$ 
  \\
  \end{tabular}
  \caption{
  A summary table for various vortices and 
  color magnetic flux tubes. The $N=3$ case corresponds to those
  in the CFL phase of dense QCD.
  NA denotes ``non-Abelian.'' 
  ``OPM'' implies the sub-OPM that vortices are supported by 
  nontrivial first homotopy groups $\pi_1$(OPM), 
  except for color flux tubes which are 
  topologically trivial: $\pi_1(SU(N)_{\rm C})=0$.
  ${\cal M}_{\rm V}$, ${\cal M}_{\rm A}$ and ${\cal M}$ are 
  the OPM for vector symmetry breaking, 
  OPM for axial and chiral symmetry breakings,
  and full OPM 
  defined in Eqs.~(\ref{eq:sub-OPM}) and (\ref{eq:fullOPM0}). 
  See Appendix \ref{sec:OPM} for details of these OPMs. 
  ``Chiral circulation'' would imply an amount of magnetic fluxes if the chiral symmetry is gauged, where the normalization is taken 
  such that a closed loop in 
  $SU(N)_{\rm L-R}$ gives a unit flux. 
  \label{tab:summary}
  }
\end{table*}
%%%%%   Table   %%%%%

%%%%%%%%%%%%%%%%%%
\section{Superfluid vortices and color magnetic flux tubes}
\label{sec:superfluid-vrtx}
In this section, 
we review superfluid vortices in the CFL phase: 
Abelian superfluid vortices 
and non-Abelian semi-superfluid vortices.

%%%%%%%%%%%%
\subsection{Abelian superfluid vortices}\label{sec:Abelian-superfluid}
The simplest vortex is 
an Abelian superfluid vortex winding around $U(1)_{\rm B}$ 
 \cite{Forbes:2001gj,Iida:2002ev}, given 
 in the polar coordinates $(r,\varphi)$ by 
\beq
\Phi_{\rm L} = - \Phi_{\rm R}  
=  e^{i\varphi} f(r) {\bf 1}_N 
= e^{i\theta_{\rm B}(\varphi)} f(r) {\bf 1}_N 
\label{eq:ASV}
\eeq
with $e^{i\theta_{\rm B}(\varphi)}=e^{i\varphi}$ 
and the profile function $f$ with the boundary conditions 
$f (r=0) = 0$ and $f (r=\infty)= v$. 
This is unstable to decay into $N$ non-Abelian semi-superfluid vortices
introduced in Sec.~\ref{sec:NAV}
\cite{Nakano:2007dr,Cipriani:2012hr,Alford:2016dco}.

In this notation of Eq.~(\ref{eq:label}), the Abelian superfluid vortex is labeled by $(N,N)$
because of $\det \Phi_{\rm L} \sim \det \Phi_{\rm R} \sim e^{Ni\varphi}$.
The gauge invariant $\Sigma$ is $\Sigma =f^2{\bf 1}_N$ having no winding.

%%%%%%%%
\subsection{Non-Abelian semi-superfluid vortices}\label{sec:NAV}
In this subsection,
we review non-Abelian semi-superfluid vortices (color magnetic flux tubes)
for comparison with chiral non-Abelian vortices 
introduced in Sec.~\ref{sec:CnAV}. 
The ansatz for a single non-Abelian semi-superfluid vortex 
winding around the sub-OPM ${\cal M}_{\rm V}\simeq U(N)_{\rm C-(L+R)+B}$ for the vector symmetry breaking 
is given 
in the polar coordinates $(r,\varphi)$ by
\beq
 && \Phi_{\rm L} = - \Phi_{\rm R} 
 = \left(
\begin{array}{cc}
f(r)e^{i\varphi} & 0 \\
0 & g(r) {\bf 1}_{N-1}
\end{array}
\right) 
=  e^{{i\over N} \varphi} 
e^{{i \over N}\varphi T_N}
 \left(
\begin{array}{cc}
f(r) & 0 \\
  0 & g(r) {\bf 1}_{N-1}
\end{array}
\right)  
\nonumber\\ && \hspace{2cm}
= e^{i\theta_{\rm B}(\varphi)} 
U(\varphi)  \left(
\begin{array}{cc}
f(r) & 0 \\
  0 & g(r) {\bf 1}_{N-1}
\end{array}
\right) 
, 
\nonumber\\
&& A_i  = - \epsilon_{ij}\frac{x^j}{N g_s r^2}(1-h(r)) T_N
\label{eq:ansatz0}
\eeq
with 
\beq 
T_N = {\rm diag.} (N-1,-1, \cdots,-1 )
\eeq
and 
\beq
 e^{i\theta_{\rm B}(\varphi)} = e^{i\varphi/N} ,\quad
 U(\varphi) = e^{{i \over N}\varphi T_N} , 
 \label{eq:vortex-trans0}
\eeq
with the boundary condition for the profile functions 
$f,g$ and $h$
\beq
(f,g',h)_{r=0} = (0,0,1),\quad
(f,g,h)_{r=\infty} = (v,v,0). \label{eq:b.c.0}
\eeq
Explicit numerical solutions were
 constructed in Ref.~\cite{Eto:2009kg}.
This carries a $1/N$ $U(1)_{\rm B}$ circulation 
compared with a unit circulation of an Abelian superfluid vortex 
given in Eq.~(\ref{eq:ASV}), 
and 
a color magnetic flux 
which is $1/N$ of that of a pure color flux tube 
generated by a closed loop in the $SU(N)_{\rm C}$ gauge group 
\cite{Iida:2004if}. The latter is unstable to decay into the ground state 
due to the trivial first homotopy group $\pi_1[SU(N)_{\rm C}]=0$.
In terms of the gauge invariants,
the non-Abelian semi-superfluid vortex is labeled by $(1,1)$
because of $\det \Phi_{\rm L} \sim \det \Phi_{\rm R} \sim e^{i\varphi}$.

More generally, 
the $SU(N)_{\rm C+L+R}$ transformation on the ansatz in Eq.~(\ref{eq:ansatz0}) yields a continuous 
family of solutions. 
They are characterized by the moduli space 
~\cite{Nakano:2007dr, Nakano:2008dc,Eto:2013hoa}
\beq
{\mathbb C}P^{N-1} = {SU(N)_{\rm C+L+R}\over SU(N-1)\times U(1)}.
\label{eq:CPN-1-0}
\eeq
These modes are normalizable \cite{Eto:2013hoa,Eto:2009tr},
and 
their effective world-sheet Lagrangian was constructed in 
a singular gauge \cite{Eto:2013hoa,Eto:2009tr} 
and a regular gauge \cite{Chatterjee:2016tml}. 
The gauge invariant $\Sigma$ is $\Sigma =\diag (f^2, g^2, \cdots, g^2)$ having no winding.
This can represent the ${\mathbb C}P^{N-1} $ orientation in Eq.~(\ref{eq:CPN-1-0}) 
at $r=0$: $\Sigma (r=0)=\diag (0, *, \cdots, *)$ with $*$ being a non-zero constant 
in the case of the orientation in Eq.~(\ref{eq:ansatz0}).
Or, we may define the orientational vector $\phi \in {\mathbb C}^N$ by $\phi \cdot \Sigma =0$, 
giving rise to $\phi^T = (*,0,\cdots,0)$ 
in the case of the orientation in Eq.~(\ref{eq:ansatz0}) 
\cite{Eto:2005yh,Eto:2006pg}.

The Abelian superfluid vortex is dynamically unstable to decay into $N$ 
non-Abelian semi-superfluid vortices \cite{Nakano:2007dr,Cipriani:2012hr,Alford:2016dco}.
This decay process can be expressed as
\beq 
 (N,N) \to N (1,1).
\eeq

%%%%%%%%%%%%%%%%%%%%%%%%%%%
\section{Abelian and non-Abelian axial vortices}
\label{sec:axial-chiral-vtx}

In this section, we discuss
Abelian 
and non-Abelian axial vortices,
which are global vortices without any color fluxes.

\subsection{Abelian axial vortices} \label{sec:Abelian-axial}
First, let us turn off 
the axial and chiral symmetry breaking terms by
$\gamma_1=\gamma_2=\gamma_3=0$. 
A single Abelian axial vortex winding around $U(1)_{\rm A}$ 
is given by
\beq
 && \Phi_{\rm L} = - \Phi_{\rm R}^\dagger  
=  e^{i\varphi} f(r) {\bf 1}_N 
= e^{i\theta_{\rm A}(\varphi)} f(r) {\bf 1}_N \label{eq:axial}
\eeq
with $e^{i\theta_{\rm A}(\varphi)}=e^{i\varphi}$ 
and the profile function $f$ with the boundary conditions 
$f (r=0) = 0$ and $f (r=\infty)= v$. 
This vortex is labeled by $(N,-N)$
because of $\det\Phi_{\rm L} \sim e^{Ni\varphi}$ and 
$\det\Phi_{\rm R} \sim e^{-Ni\varphi}$.

In the presence of 
the axial and chiral symmetry breaking terms,
$\gamma_{1,2,3} \neq 0$, domain walls are attached to the vortex. 
To see this, we consider an infinitely large circle 
with the spatial angle $\varphi$ 
encircling 
the axial vortex.
Then, 
let us substitute the ansatz in Eq.~(\ref{eq:axial}) to
 the potential $V$ in Eq.~(\ref{eq:GL}), 
 with replacing the spatial angle $\varphi$ 
by a function $\phi(\varphi)$ 
depending on the angle $\varphi$ 
with the boundary condition
$\phi(\varphi=0) =0$ 
and $\phi(\varphi=2\pi) = 2\pi$.   
Then, the potential can be evaluated at spatial infinities as
\beq
 V = 2 N \gamma_1 \cos (2 \phi(\varphi)) 
 + 2 N \gamma_2 \cos (4\phi(\varphi)) 
 + 2 \gamma_3 \cos (2N \phi(\varphi) ).\label{eq:pot-abelian-axial}
\eeq 
Along the large circle at infinity encircling the axial vortex, 
there is also the gradient term. Thus, 
the effective energy for $\phi$ at the large circle becomes  
${\cal E}_{\rm eff} = N v^2 (\partial_{\varphi} \phi)^2 + V$. 
This is a variant of an $N$-ple sine-Gordon model.

In the case of
$(\gamma_1,\gamma_2,\gamma_3) 
=(0,0,\gamma_3)$,
the Abelian axial vortex is attached by 
$2N$ domain walls.
In this case, 
this vortex is unstable to decay into $N$ 
non-Abelian axial vortices 
introduced in Sec.~\ref{sec:NAGV}, each of which is attached by two domain walls. 
See discussion in Sec.~\ref{sec:decay}.

If we turn on $\gamma_{1,2}$, 
these $2N$ domain walls would constitute a composite wall in general.
It is an open question whether the decay is suppressed or not 
in such a case.

Before closing this subsection, let us mention a relation 
to analogous axial vortices in 
the context of chiral symmetry breaking at low density.
In that case, 
the axial vortex is attached by $N$ domain walls  
\cite{Balachandran:2001qn}, 
and it decays into non-Abelian global strings 
each of which is attached by one domain wall
\cite{Eto:2013hoa,Eto:2013bxa}. 
To compare these two cases, 
it is convenient to see the gauge invariant $\Sigma$ in Eq.~(\ref{eq:Sigma}). 
In terms of this gauge invariant, 
the axial vortex in Eq.~(\ref{eq:axial}) can be rewritten as
\beq
 \Sigma  
= - e^{2i\varphi} f^2(r) {\bf 1}_N 
= -e^{2i\theta_{\rm A}(\varphi)} f^2(r) {\bf 1}_N. \label{eq:axial2}
\eeq
Thus, one can see that 
the minimum winding of the axial vortex in the CFL phase 
corresponds to the double winding of the axial vortex at low density 
\cite{Balachandran:2001qn,Eto:2013hoa,Eto:2013bxa}.

%%%%%%%%%
\subsection{Non-Abelian axial vortices}
\label{sec:NAGV}
Here we discuss non-Abelian axial vortices 
winding in 
the sub-OPM ${\cal M}_{\rm A} = U(N)_{\rm L-R+A}$ for the axial and chiral symmetry breakings.
An analogue of this at low density was 
discussed in linear sigma models in Refs.~\cite{Balachandran:2002je,Nitta:2007dp,Nakano:2007dq,Nakano:2008dc,Eto:2009wu}. 
First, let us turn off 
the axial and chiral symmetry breaking terms by
$\gamma_1=\gamma_2=\gamma_3=0$. 

The ansatz for a single non-Abelian axial vortex 
 is given in the polar coordinates $(r,\varphi)$ by
\beq
 \Phi_{\rm L} &=& \left(
\begin{array}{cc}
 f(r)e^{i\varphi} & 0 \\
                   0 & g(r) {\bf 1}_{N-1}
\end{array}
\right) 
=  e^{{i\over N} \varphi}
 \left(
\begin{array}{cc}
f(r) & 0 \\
  0 & g(r) {\bf 1}_{N-1}
\end{array}
\right)  e^{{i \over N}\varphi T_N} \nonumber\\
&& \hspace{3.7cm}= e^{i\theta_{\rm A}(\varphi)} 
\left(
\begin{array}{cc}
f(r) & 0 \\
  0 & g(r) {\bf 1}_{N-1}
\end{array}
\right) U(\varphi), 
\nonumber\\
 - \Phi_{\rm R}  &=&  
\left(\begin{array}{cc}
f(r)e^{- i\varphi} & 0 \\
0 & g(r) {\bf 1}_{N-1}
\end{array}
\right) 
=  e^{-{i\over N} \varphi}  
 \left(
\begin{array}{cc}
f(r) & 0 \\
  0 & g(r) {\bf 1}_{N-1}
\end{array}
\right)  e^{-{i \over N}\varphi T_N} \nonumber\\
&& \hspace{3.7cm}= e^{-i\theta_{\rm A}(\varphi)} 
\left(
\begin{array}{cc}
f(r) & 0 \\
  0 & g(r) {\bf 1}_{N-1}
\end{array}
\right) U^\dagger(\varphi)
, 
\label{eq:ansatz02}
\eeq
with 
\beq
 e^{i\theta_{\rm A}(\varphi)} = e^{i\varphi/N}, \quad
 U(\varphi) = e^{{i \over N}\varphi T_N} . 
 \label{eq:vortex-trans02}
\eeq
The boundary condition is 
\beq
(f,g',h)_{r=0} = (0,0,1),\quad
(f,g,h)_{r=\infty} = (v,v,0). \label{eq:b.c.1}
\eeq
Explicit numerical solutions can be found in Refs.~\cite{Nitta:2007dp,
Eto:2009wu}.
This vortex is a purely global vortex without any color magnetic flux, 
carrying a $U(1)_{\rm A}$ winding number which is $1/N$ of that of 
the Abelian axial vortex in Eq.~(\ref{eq:axial}).  
The non-Abelian axial vortex is labeled by $(1,-1)$
because of $\det \Phi_{\rm L} \sim e^{i\varphi}$ and $\det \Phi_{\rm R} \sim e^{-i\varphi}$.
 
A set of solutions has ${\mathbb C}P^{N-1}$ moduli, 
which are non-normalizable  
since the $SU(N)_{\rm L+R}$ transformation changes 
the boundary.

In the presence of 
the axial and chiral symmetry breaking terms,
$\gamma_{1,2,3} \neq 0$, domain walls are attached to the vortex.
In order to understand domain walls attached 
to the non-Abelian axial vortex, 
let us substitute the ansatz in Eq.~(\ref{eq:ansatz02}) to
 the potential $V$ in (\ref{eq:GL}), 
 with replacing the spatial angle $\varphi$ 
by a function $\phi(\varphi)$ 
depending on the angle $\varphi$ 
with the boundary condition
$\phi(\varphi=0) =0$ 
and $\phi(\varphi=2\pi) = 2\pi$.   
Then, the potential can be evaluated at spatial infinities as
\beq
 V = 2 (\gamma_1+ \gamma_3) \cos (2 \phi(\varphi)) 
 + 2 \gamma_2 \cos (4\phi(\varphi)) .\label{eq:DSG-half}
\eeq
Together with the gradient term, 
the effective energy for $\phi$ on the large circle 
at infinity encircling the non-Abelian axial vortex 
 becomes  
${\cal E}_{\rm eff} = v^2 (\partial_{\varphi} \phi)^2 + V$. 
This is the double sine-Gordon model 
with a half periodicity $\pi$ instead of the usual case of $2\pi$.

In the case of 
$(\gamma_1,\gamma_2,\gamma_3) 
=(\gamma_1,0,\gamma_3)$,
one non-Abelian axial vortex is attached by 
two domain walls. 
These two domain walls are attached 
from the opposite sides of the vortex. 
This vortex is unstable to decay into 
two chiral non-Abelian vortices 
introduced in Sec.~\ref{sec:chiralNA},
each of which is attached by one domain wall.\footnote{
In the context of chiral symmetry breaking at low density,
the axial vortex is attached by one domain wall  
\cite{Balachandran:2002je}, and decays do not occur.
}
See discussion in Sec.~\ref{sec:decay}.

If we turn on $\gamma_2$, 
how these two domain walls attach to the vortex depends on the parameters 
$\gamma_1, \gamma_2$
as classified in Refs.~\cite{Eto:2018hhg,Eto:2018tnk} 
in the context of two-Higgs doublet models.
In some case, 
these two domain walls constitute a composite wall.
It is an open question whether the decay is suppressed or not in this case.

In terms of the gauge invariant $\Sigma$ in Eq.~(\ref{eq:Sigma}), 
the ansatz in Eq.~(\ref{eq:ansatz02}) can be rewritten as
\beq
&&
-\Sigma = 
 \left(
\begin{array}{cc}
F(r)e^{2 i\varphi} & 0 \\
0 & G(r) {\bf 1}_{N-1}
\end{array}
\right) 
=  e^{{2i \over N} \varphi} 
  e^{{i \over N}\varphi T_N} 
 \left(
\begin{array}{cc}
F(r) & 0 \\
  0 & G(r) {\bf 1}_{N-1}
\end{array}
\right)  e^{{i \over N}\varphi T_N} \nonumber\\
&& \hspace{5.2cm}= 
e^{2 i\theta_{\rm A}(\varphi)} 
U(\varphi)  
\left(
\begin{array}{cc}
F(r) & 0 \\
  0 & G(r) {\bf 1}_{N-1}
\end{array}
\right) U(\varphi)
,  
\label{eq:ansatz02-sigma}
\eeq
with $F\equiv f^2$ and $G=g^2$, and $U$ and $\theta_{\rm A}$ in Eq.~(\ref{eq:vortex-trans02}).
It is obvious that this vortex has a double-winding 
compared with the corresponding one 
in the linear sigma model.
In fact, the one with unit winding in $\Sigma$ 
discussed in Sec.~10 of the review paper \cite{Eto:2013hoa}
 corresponds to 
the chiral non-Abelian vortex 
introduced in the next section.
  
%%%%%%%%%%%

\section{Chiral non-Abelian vortices}\label{sec:chiralNA}
\label{sec:CnAV}

In this section, we introduce a novel vortex of non-Abelian kind, 
that is, chiral non-Abelian vortices.
Here, we restrict ourselves to the case 
in the absence of the chiral symmetry breaking terms: 
$\gamma_1=\gamma_2=\gamma_3=0$ 
in which the axial and chiral symmetries become exact.
We then discuss the topological obstruction 
and AB phases around these vortices.

%%%%%%%%%%%%
\subsection{Solutions of chiral non-Abelian vortices}
\begin{figure}
\begin{center}
\begin{tabular}{cc}
\includegraphics[width=7cm]{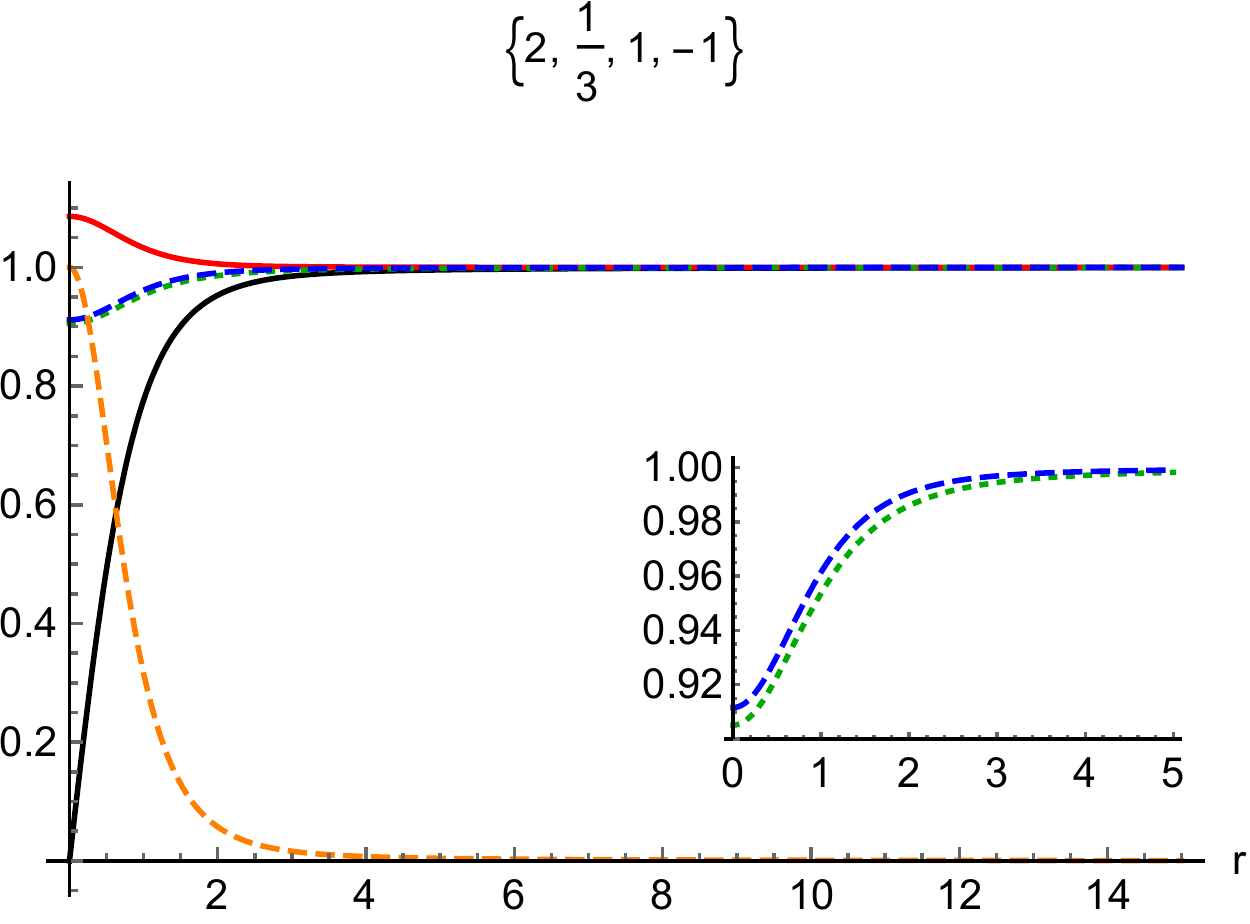}&
\includegraphics[width=7cm]{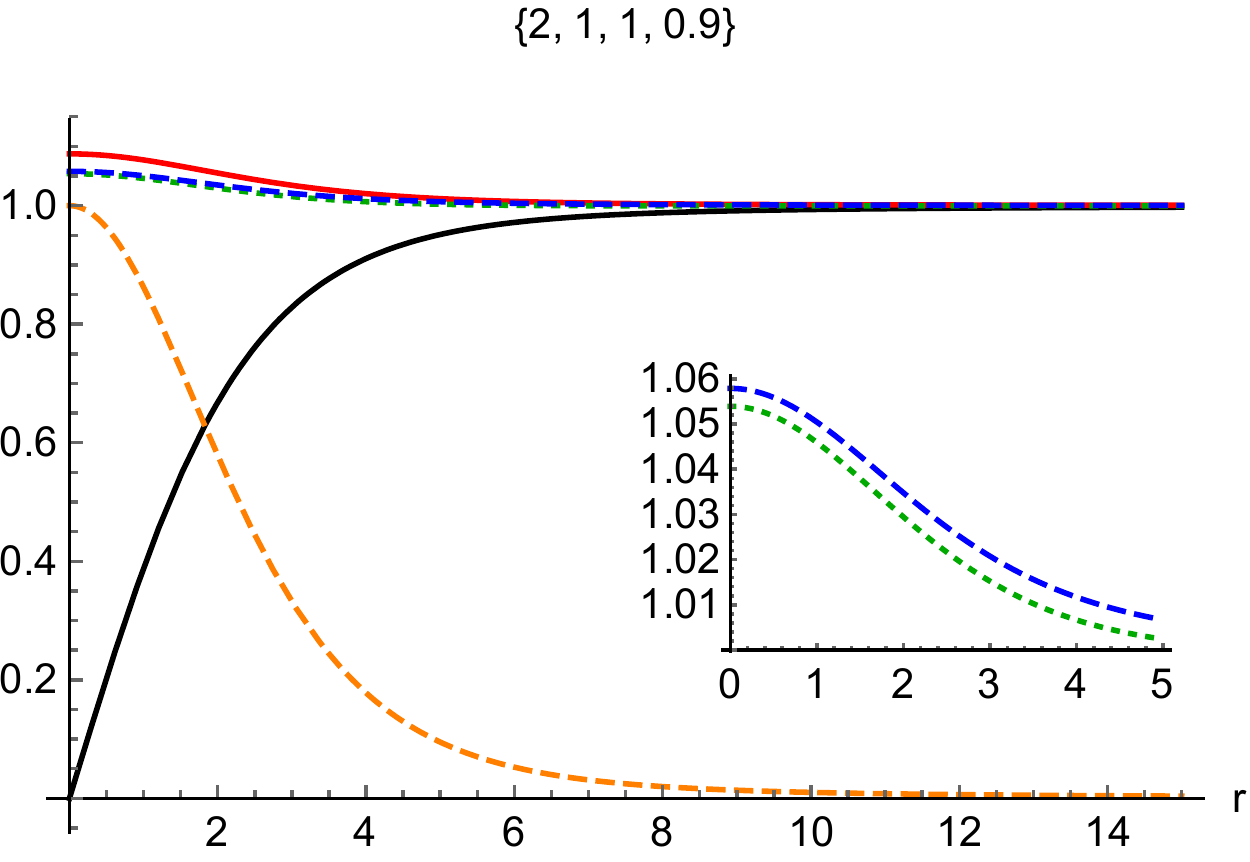}\\\ & \\
\includegraphics[width=7cm]{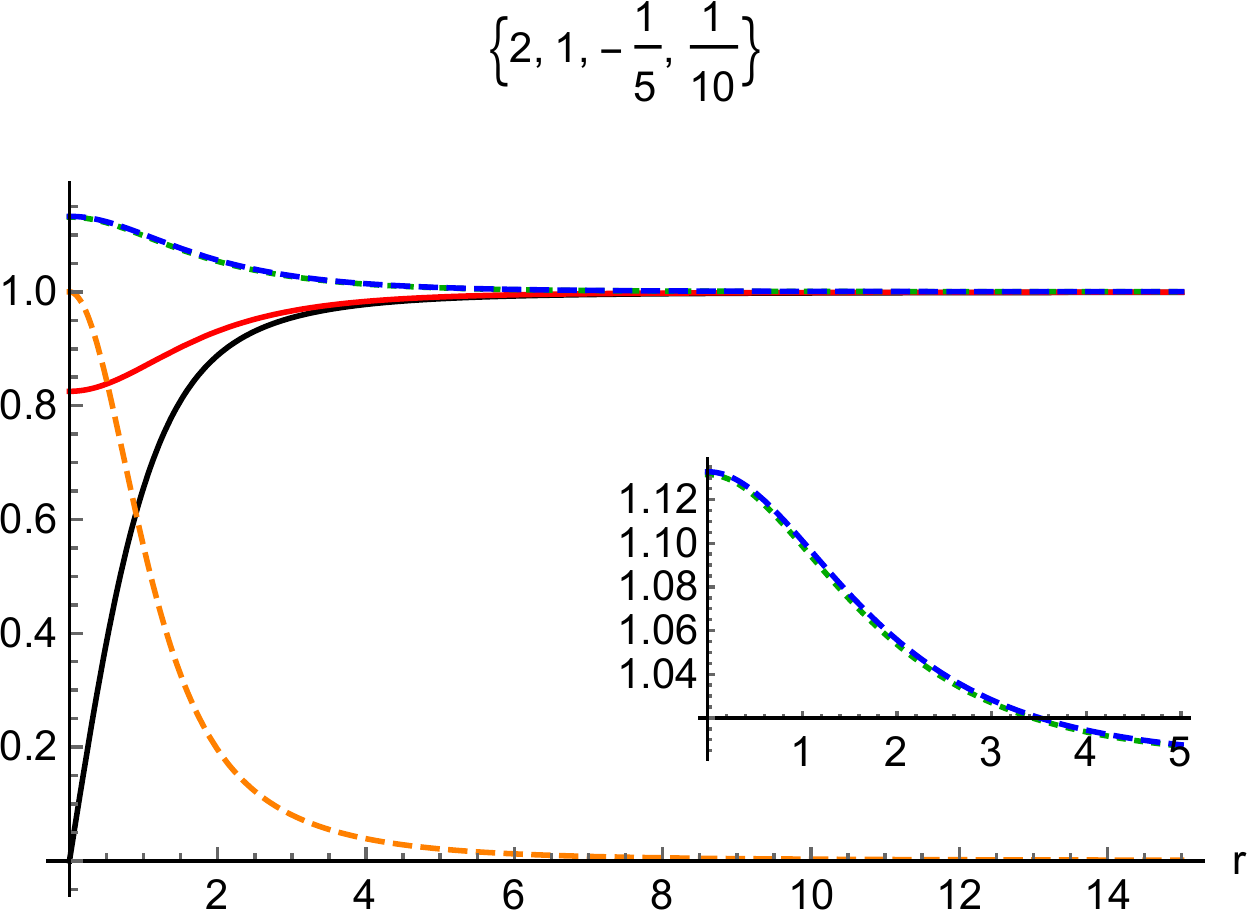}&
\includegraphics[width=7cm]{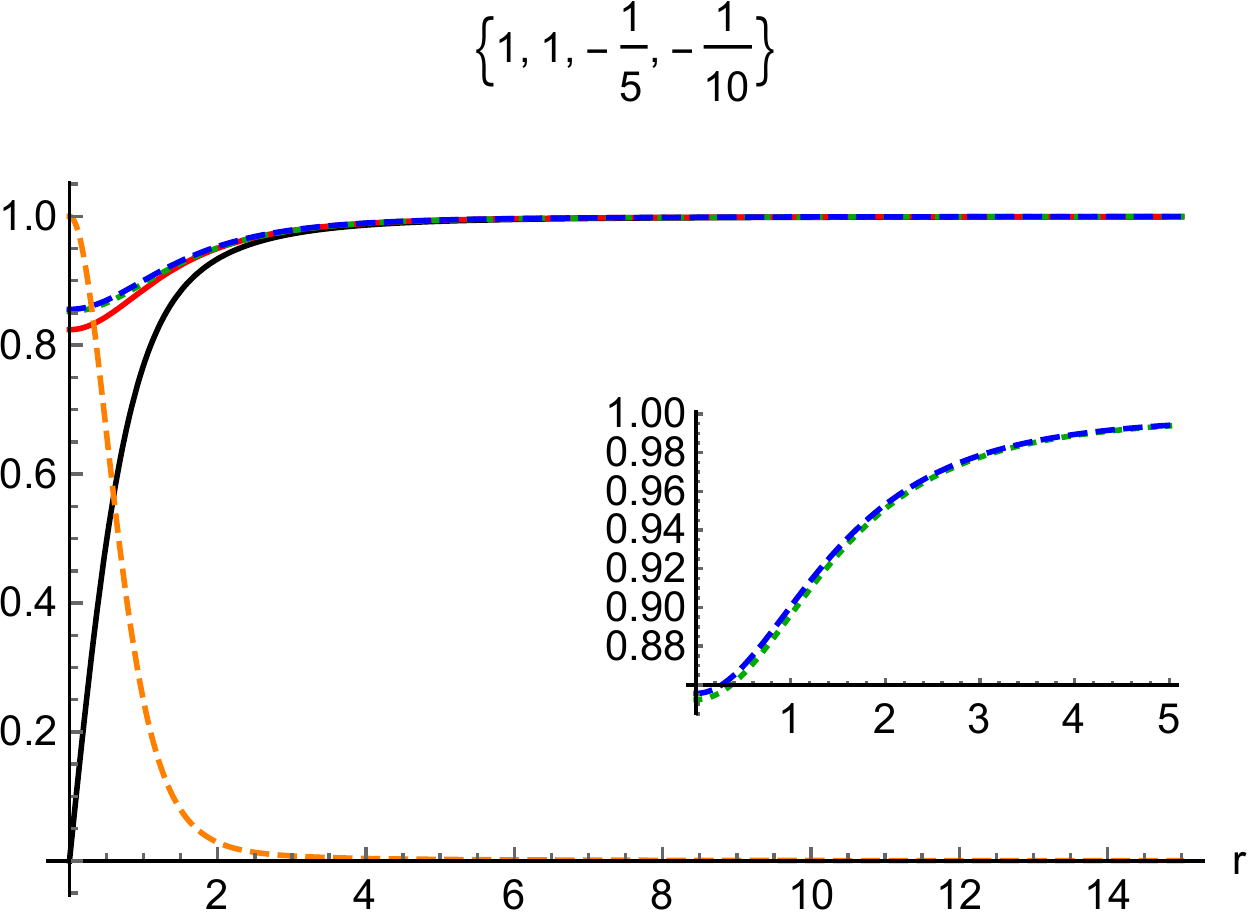}
\end{tabular}
\end{center}
\caption{Numerical solutions of a single left (right) chiral non-Abelian vortex for $N=3$.
The black-solid, red-solid, green-dotted, blue-dashed, and orange-dashed lines correspond to
$f(r)$, $g(r)$, $c(r)$, $d(r)$, and $h(r)$, respectively. The label at top of each
panel shows the parameter combination $\{m,\lambda_1,\lambda_2,\lambda_3\}$. 
The profiles $c(r)$ and $d(r)$ are almost degenerate for all the cases, and
the insets show small deviation between $c(r)$ and $d(r)$ near the origin.
}
\label{fig:N=3}
\end{figure}

Chiral non-Abelian vortices introduced in this section 
are the minimum vortices in the CFL phase. 
There are two kinds of chiral non-Abelian vortices, 
namely of left and right chiralities, given by 
\beq
  \mbox{Left } (1,0): &&\det  \Phi_{\rm L} \sim e^{i\varphi}, \quad \det  \Phi_{\rm R} \sim 1,  \non
 \mbox{Right } (0,1): &&\det  \Phi_{\rm L} \sim 1, \hspace{7mm} \det  \Phi_{\rm R} \sim e^{i\varphi},
\eeq 
respectively.
In order to construct these configurations,
we note the relations for labels 
\beq
 && (1,0) = {1\over 2} \left[(1,1) + (1,-1)\right],\non
 && (0,1) = {1\over 2} \left[(1,1) - (1,-1)\right].
\eeq
These imply that a chiral non-Abelian vortex 
can be constructed as a sum of 
a half non-Abelian semi-superfluid vortex 
and a half non-Abelian axial vortex. 
We thus reach the ansatz for a chiral non-Abelian vortex 
of the left chirality $(1,0)$,  given 
 in the polar coordinates $(r,\varphi)$
 by
\beq
 \Phi_{\rm L} &=& \left(
\begin{array}{cc}
f(r)e^{i\varphi} & 0 \\
0 & g(r) {\bf 1}_{N-1}
\end{array}
\right) 
=  e^{{i\over 2N} \varphi}    e^{{i\over 2N} \varphi} 
e^{{i \over 2N}\varphi T_N}
 \left(
\begin{array}{cc}
f(r) & 0 \\
  0 & g(r) {\bf 1}_{N-1}
\end{array}
\right)  e^{{i \over 2N}\varphi T_N} \nonumber\\
&& \hspace{3.7cm}= e^{i\theta_{\rm B}(\varphi)+i\theta_{\rm A}(\varphi)} 
U(\varphi)  \left(
\begin{array}{cc}
f(r) & 0 \\
  0 & g(r) {\bf 1}_{N-1}
\end{array}
\right) U(\varphi)
,
\nonumber\\
 - \Phi_{\rm R}  &=&  \left(
\begin{array}{cc}
c(r) & 0 \\
   0 & d(r) {\bf 1}_{N-1}
\end{array}
\right)
= e^{{i\over 2N} \varphi}   e^{-{i\over 2N} \varphi}  e^{{i \over 2N}\varphi T_N}
 \left(
\begin{array}{cc}
c(r) & 0 \\
   0 & d(r) {\bf 1}_{N-1}
\end{array}
\right)  e^{-{i \over 2N}\varphi T_N}\nonumber\\
&& \hspace{3.3cm} 
= e^{i\theta_{\rm B}(\varphi)-i\theta_{\rm A}(\varphi)} 
U(\varphi)  \left(
\begin{array}{cc}
c(r) & 0 \\
   0 & d(r) {\bf 1}_{N-1}
\end{array} 
\right) U^\dagger(\varphi) \nonumber\\
 A_i  &=& - \epsilon_{ij}\frac{x^j}{2N g_s r^2}(1-h(r)) T_N
\label{eq:ansatz}
\eeq
with 
\beq
 e^{i\theta_{\rm B}(\varphi)} = e^{i\varphi/2N} ,\quad
 e^{i\theta_{\rm A}(\varphi)} = e^{i\varphi/2N}, \quad
 (g_{\rm C} = U_{\rm L}^\dagger = U_{\rm R}=) U(\varphi) = e^{{i \over 2N}\varphi T_N} . 
 \label{eq:vortex-trans}
\eeq
The equations of motions for the profile functions are given by
\beq
&&f''+\frac{f'}{r} - \frac{ ((N-1)h+(N+1))^2}{4N^2 r^2}f +\frac{m^2}{2}  f\non
&&\quad-\frac{1}{2} \left[\left(\lambda _1+\lambda _2\right) f^2+ (N-1) \lambda _2 g^2 + \lambda _3 c^2+ (N-1) \lambda _3d^2\right] f = 0,\\
&&g''+\frac{g'}{r}-\frac{(h-1)^2}{4N^2 r^2}g +\frac{m^2}{2}  g \non
&&\quad -\frac{1}{2}  \left[\lambda _2 f^2+\left(\lambda _1+(N-1) \lambda _2\right) g^2 + \lambda _3 c^2+(N-1) \lambda _3d^2\right]g = 0,\\
&&c''+\frac{c'}{r}-\frac{(N-1)^2(h-1)^2}{4N^2 r^2}c + \frac{m^2}{2} c \non
&&\quad -\frac{1}{2} \left[\lambda _3f^2+ (N-1) \lambda _3g^2 + \left(\lambda _1+\lambda _2\right) c^2+ (N-1) \lambda _2 d^2\right]c = 0,\\
&&d''+\frac{d'}{r}-\frac{(h-1)^2}{4N^2 r^2}d+\frac{m^2}{2} d \non
&&\quad-\frac{1}{2} \left[\lambda _3 f^2+ (N-1) \lambda_3g^2 + \lambda _2 c^2+\left(\lambda _1+ (N-1) \lambda _2\right) d^2\right] d = 0,\\
&&h''-\frac{h'}{r} -\frac{2g_s^2}{N} \big[\left((N-1) f^2+g^2 + (N-1) c^2+d^2\right) h \non
&&\quad  + (N+1) f^2-g^2 - (N-1) c^2-d^2\big] = 0.
\eeq
The boundary condition is 
\beq
(f,g',c',d',h)_{r=0} = (0,0,0,0,1),\quad
(f,g,c,d,h)_{r=\infty} = (v,v,v,v,0). \label{eq:b.c.}
\eeq
Numerical solutions for several typical parameter combinations in the case of $N=3$
are plotted in Fig.~\ref{fig:N=3}.

Compared with the ansatz 
for the usual non-Abelian semi-superfluid vortex in Eq.~(\ref{eq:ansatz0}), 
 the $U(1)_{\rm B}$ and $SU(3)_{\rm C}$ actions are halves of those 
of  the ansatz in Eq.~(\ref{eq:ansatz0}) 
winding in the sub-OPM ${\cal M}_{\rm V} = U(N)_{\rm C-(L+R)+B}$, 
 and the rests are complemented by 
 going through 
the other sub-OPM ${\cal M}_{\rm A} = U(N)_{\rm L-R+A}$ 
 for the axial and chiral symmetry breakings,
 which are halves of 
 non-Abelian axial vortices
 in Eq.~(\ref{eq:ansatz02}).
A closed loop surrounding the chiral non-Abelian vortex is mapped onto a closed loop in the full
OPM ${\cal M}$ given in Eq.~(\ref{eq:fullOPM0}), 
and consequently, this carries $1/2N$ $U(1)_{\rm B}$ circulation and 
the color magnetic flux, 
both of which are halves of those of the usual non-Abelian semi-superfluid vortex. 
The color magnetic flux is $1/2N$ of that of a pure color flux tube.

In the above ansatz in Eq.~(\ref{eq:ansatz}), 
we have considered the winding in the (1,1) component.
Instead, we can embed it in other diagonal components, 
thus finding $N$ solutions of the same energy, 
as the case of the usual non-Abelian semi-superfluid vortex.
Again, more generally, 
the $SU(N)_{\rm C+L+R}$ transformation on the ansatz in Eq.~(\ref{eq:ansatz}) yields a continuous 
family of solutions, again characterized by the moduli space 
\beq
{\mathbb C}P^{N-1} = {SU(N)_{\rm C+L+R}\over SU(N-1)\times U(1)}.
\label{eq:CPN-1}
\eeq

Likewise, we also can construct  a vortex  
of the right chirality $(0,1)$ winding in $\Phi_{\rm R}$ in the same way:
\beq
  \Phi_{\rm L} & =&  \left(
\begin{array}{cc}
c(r) & 0 \\
   0 & d(r) {\bf 1}_{N-1}
\end{array}
\right)
= e^{{i\over 2N} \varphi}   e^{-{i\over 2N} \varphi}  e^{{i \over 2N}\varphi T_N}
 \left(
\begin{array}{cc}
c(r) & 0 \\
   0 & d(r) {\bf 1}_{N-1}
\end{array}
\right)  e^{-{i \over 2N}\varphi T_N}\nonumber\\
&& \hspace{3.2cm} = e^{i\theta_{\rm B}(\varphi)+i\theta_{\rm A}(\varphi)} 
U(\varphi)  \left(
\begin{array}{cc}
c(r) & 0 \\
   0 & d(r) {\bf 1}_{N-1}
\end{array} 
\right) U^\dagger(\varphi) \nonumber\\
- \Phi_{\rm R} &=& \left(
\begin{array}{cc}
f(r)e^{i\varphi} & 0 \\
0 & g(r) {\bf 1}_{N-1}
\end{array}
\right) 
=  e^{{i\over 2N} \varphi}    e^{{i\over 2N} \varphi} 
e^{{i \over 2N}\varphi T_N}
 \left(
\begin{array}{cc}
f(r) & 0 \\
  0 & g(r) {\bf 1}_{N-1}
\end{array}
\right)  e^{{i \over 2N}\varphi T_N} \nonumber\\
&& \hspace{3.8cm}= e^{i\theta_{\rm B}(\varphi)-i\theta_{\rm A}(\varphi)} 
U(\varphi)  \left(
\begin{array}{cc}
f(r) & 0 \\
  0 & g(r) {\bf 1}_{N-1}
\end{array}
\right) U(\varphi)
,
\nonumber\\
 A_i  &=& - \epsilon_{ij}\frac{x^j}{2N g_s r^2}(1-h(r)) T_N,
\label{eq:ansatz2}
\eeq
with the same profile functions as those 
in Eq.~(\ref{eq:ansatz}) 
and the same boundary conditions for them 
with Eq.~(\ref{eq:b.c.}).
This carries 
the same $U(1)_{\rm B}$ circulation
and the {\it same} color magnetic flux with those 
with the left one in Eq.~(\ref{eq:ansatz}),
but the $U(1)_{\rm A}$ and $SU(N)_{\rm L-R}$ transformations 
are opposite to those of the left one in Eq.~(\ref{eq:ansatz}):
\beq
 e^{i\theta_{\rm B}(\varphi)} = e^{i\varphi/2N} ,\quad
 e^{i\theta_{\rm A}(\varphi)} = e^{- i\varphi/2N}, \quad
 (g_{\rm C} = U_{\rm L} = U_{\rm R}^\dagger=) U(\varphi) = e^{{i \over 2N}\varphi T_N} . 
 \label{eq:vortex-trans2}
\eeq
A continuous family of solutions is parametrized by a copy of the moduli space in Eq.~(\ref{eq:CPN-1}).

In terms of the gauge invariant $\Sigma$ in Eq.~(\ref{eq:Sigma}), 
the chiral non-Abelian vortex
of the left chirality $(1,0)$ in Eq.~(\ref{eq:ansatz}) 
can be rewritten as 
\beq
&&
-\Sigma = 
 \left(
\begin{array}{cc}
 F(r) e^{+ i\varphi} & 0 \\
                          0 & G(r) {\bf 1}_{N-1}
\end{array}
\right) 
=  e^{{i \over N} \varphi} 
  e^{-{i \over 2N}\varphi T_N} 
 \left(
\begin{array}{cc}
 F(r) & 0 \\
        0 & G(r)  {\bf 1}_{N-1}
\end{array}
\right)  e^{-{i \over 2N}\varphi T_N} \nonumber\\
&& \hspace{5.2cm}= 
e^{2 i\theta_{\rm A}(\varphi)} 
U(\varphi)  
\left(
\begin{array}{cc}
 F(r) & 0 \\
    0 & G(r) {\bf 1}_{N-1}
\end{array}
\right) U(\varphi)
,  
\eeq
with $F\equiv fc$, $G\equiv gd$, and 
$U(\varphi)$ and $e^{2 i\theta_{\rm A}(\varphi)}$ in Eq.~(\ref{eq:vortex-trans}), 
while 
the one of the right chirality $(0,1)$ in Eq.~(\ref{eq:ansatz2}) 
can be rewritten as 
\beq
&&
-\Sigma = 
 \left(
\begin{array}{cc}
 F(r) e^{- i\varphi} & 0 \\
                       0 & G(r) {\bf 1}_{N-1}
\end{array}
\right) 
=  e^{-{i \over N} \varphi} 
  e^{-{i \over 2N}\varphi T_N} 
 \left(
\begin{array}{cc}
 F(r) & 0 \\
    0 & G(r) {\bf 1}_{N-1}
\end{array}
\right)  e^{-{i \over 2N}\varphi T_N} \nonumber\\
&& \hspace{5.3cm}= 
e^{2 i\theta_{\rm A}(\varphi)} 
U^\dagger(\varphi)  
\left(
\begin{array}{cc}
 F(r) & 0 \\
    0 & G(r) {\bf 1}_{N-1}
\end{array}
\right) U^\dagger(\varphi)
,  
\label{eq:ansatz1-2-sigma}
\eeq
with 
$F\equiv fc$, $G\equiv gd$, and 
$U(\varphi)$ and $e^{2 i\theta_{\rm A}(\varphi)}$ in Eq.~(\ref{eq:vortex-trans2}).
These two just look like an anti-vortex to each other.
In other words, 
vortices labeled by 
$(1,0)$ and $(0,-1)$ have the same form of $\Sigma$.
However, the vortices $(1,0)$ and $(0,-1)$ are distinct 
because the color magnetic fluxes that they carry 
are opposite to each other, 
which are invisible in $\Sigma$.

With this regards,
vortices in the linear sigma model in terms of $\Sigma$ 
discussed in Sec.~10 of the review paper \cite{Eto:2013hoa} 
actually describe chiral non-Abelian vortices discussed 
in this section and should 
carry magnetic fluxes (invisible in the linear sigma model), 
although this fact was not recognized in Ref.~\cite{Eto:2013hoa}.

%%%%
\subsection{Topological obstruction}
Here we discuss the so-called topological obstruction 
\changed{
(see Appendix \ref{sec:topological-obstruction} for its definition)}  
common for the vortices with the left and right chiralities 
 in Eqs.~(\ref{eq:ansatz}) and (\ref{eq:ansatz2}).
If we encircle the vortex, 
the generators $T_A$ ($A=1,\cdots,N^2-1$) of the $SU(N)_{\rm C}$ gauge group 
transform accordingly as
\beq
 T_A(\varphi) &\equiv& U (\varphi)T_A U^\dagger(\varphi)\non
 &=& \exp\left[ {i\varphi \over 2N} {\rm diag.} (N-1,-1, \cdots,-1 )\right]
T_A 
\exp\left[ -{i\varphi \over 2N} {\rm diag.} (N-1,-1, \cdots,-1 )\right]
\non
&=& 
\left(
\begin{array}{c|c}
                     (T_A)_{ij} & e^{+i\varphi/2} (T_A)_{1j} \\ \hline
e^{-i\varphi/2} (T_A)_{i1} & (T_A)_{ij}
\end{array}
\right) 
\eeq
with $i,j = 2, \cdots, N$.
After complete encirclement ($\varphi=2\pi$), these become
\beq
 T_A(\varphi=2\pi) 
=
\left(
\begin{array}{c|c}
   (T_A)_{11} & - (T_A)_{1j} \\ \hline
 - (T_A)_{i1} & (T_A)_{ij}
\end{array}
\right) \neq  T_A(\varphi=0) 
\eeq
implying that the off-diagonal blocks 
are not single-valued around the vortex.
Those off-diagonal blocks correspond to 
the broken generators of the ${\mathbb C}P^{N-1}$ moduli in Eq.~(\ref{eq:CPN-1}). 
This phenomenon is known as the topological obstruction.
More precisely, the obstruction is present for 
the CFL symmetry in Eq.~(\ref{eq:CFL}) rather than the gauge symmetry 
itself. 

The two complete encirclements give
\beq
 T_A(\varphi=4\pi) 
=T_A(\varphi=0) .
\eeq
This also implies that there is no obstruction around the 
usual non-Abelian semi-superfluid vortex  in Eq.~(\ref{eq:ansatz0}).

For vortices with different color magnetic 
fluxes corresponding to 
the ${\mathbb C}P^{N-1}$ moduli in Eq.~(\ref{eq:CPN-1}), 
corresponding broken generators have the obstruction.

%%%%%%%%%%%%%
\subsection{Generalized Aharonov-Bohm phases}
\label{sec:gAB}
AB phases around the usual non-Abelian semi-superfluid vortices 
were studied for the electromagnetism \cite{Chatterjee:2015lbf},
and for color gauge field \cite{Cherman:2018jir,
Chatterjee:2018nxe,Chatterjee:2019tbz}. 
Here, we do not consider electromagnetism. 
Let us discuss generalized AB phases around a single 
chiral non-Abelian vortex. 
In the CFL phase ($N=3$), the light quarks $q$ and heavy quarks 
$Q$ receive the following 
transformations from Eqs.~(\ref{eq:ansatz}) 
and (\ref{eq:vortex-trans}) when they encircle 
the vortex.

The heavy quarks not participating condensations receive 
ordinary AB phases contributed only from the gauge symmetry as
\beq
 Q_{\rm L} &\to& 
g_{\rm C}^*(\varphi) Q_{\rm L}
=  
 \exp\left[- {i\varphi \over 2N} {\rm diag.} (N-1,-1, \cdots,-1 )\right]
Q_{\rm L} ,\non
 Q_{\rm R} &\to& 
g_{\rm C}^*(\varphi) Q_{\rm R} 
= 
 \exp\left[- {i\varphi \over 2N} {\rm diag.} (N-1,-1, \cdots,-1 )\right] 
 Q_{\rm R}.
\eeq
After complete encirclement ($\varphi=2\pi$), these phases become
\beq
 Q_{\rm L} &\to&
 \exp\left[- {i \pi \over N} {\rm diag.} (N-1,-1, \cdots,-1 )\right]
Q_{\rm L} 
= {\rm diag.} (\epsilon^{-N+1},\epsilon, \cdots,\epsilon) Q_{\rm L}
,\non
 Q_{\rm R} &\to& 
 \exp\left[- {i \pi \over N} {\rm diag.} (N-1,-1, \cdots,-1 )\right] 
 Q_{\rm R} 
 ={\rm diag.} (\epsilon^{-N+1},\epsilon, \cdots,\epsilon) Q_{\rm R},
 \label{eq:AB-phase-Q}
\eeq
with $\epsilon$ is the $2N$-th root of the unity, 
\beq 
\epsilon = \exp (\pi i/N), \quad (\epsilon^{2N}=1).
\eeq
These form a ${\mathbb Z}_{2N}$ group.
This is a color non-singlet, 
implying that heavy quarks can detect the color of the magnetic flux of the vortex 
from infinite distances.
Note that, after two successive encirclements ($\varphi=4\pi$), they become
\beq
 Q_{\rm L} &\to& 
{\rm diag.} (\epsilon^{-2N+2},\epsilon^2, \cdots,\epsilon^2) Q_{\rm L}
 = \epsilon^2 Q_{\rm L}
,\non
 Q_{\rm R} &\to&  
 {\rm diag.} (\epsilon^{-2N+2},\epsilon^2, \cdots,\epsilon^2) Q_{\rm R}
  =  \epsilon^2 Q_{\rm R}. \label{eq:AB-phase-Q-twice}
\eeq
Thus, even numbers of manipulations give 
a ${\mathbb Z}_N$ group, 
which is a color singlet.

On the other hand, 
the light quarks participate condensations,
thus receiving generalized AB phases 
consisting of two contributions from the vortex winding
and AB phases purely coming from the color gauge group,
as was studied for usual non-Abelian semi-superfluid vortices 
in the CFL phase 
\cite{Chatterjee:2018nxe,Chatterjee:2019tbz} 
as well as non-Abelian Alice strings in the 2SC+$\dd$ phase \cite{Fujimoto:2020dsa,Fujimoto:2021bes,Fujimoto:2021wsr} (see Appendix \ref{sec:two-flavors}). 
In our case, generalized AB phases around a chiral non-Abelian vortex are  
\beq
 q_{\rm L} &\to& e^{i\theta_{\rm B}(\varphi)/2} e^{i\theta_{\rm A}(\varphi)/2} g_{\rm C}^*(\varphi) q_{\rm L} U_{\rm L}^T(\varphi)\non
&=& e^{i\varphi/2N} 
 \exp\left[- {i\varphi \over 2N} {\rm diag.} (N-1,-1, \cdots,-1 )\right]
q_{\rm L} 
\exp\left[ -{i\varphi \over 2N} {\rm diag.} (N-1,-1, \cdots,-1 )\right]
 \non
 &=&
\left(
\begin{array}{c|c}
 e^{-i ((2N-3)/2N) \varphi} (q_{\rm L})_{11} & e^{-i ((N-3)/2N) \varphi}  (q_{\rm L})_{1j}  \\ \hline
e^{ -i((N-3)/2N) \varphi}  (q_{\rm L})_{i1}  & e^{i (3/2N) \varphi}  (q_{\rm L})_{ij} 
\end{array}
\right) ,\non
 q_{\rm R} &\to& e^{i\theta_{\rm B}(\varphi)/2} e^{-i\theta_{\rm A}(\varphi)/2} g_{\rm C}^*(\varphi) q_{\rm R} U_{\rm R}^T(\varphi) \non
&=& 
 \exp\left[- {i\varphi \over 2N} {\rm diag.} (N-1,-1, \cdots,-1 )\right]
q_{\rm R} 
\exp\left[ +{i\varphi \over 2N} {\rm diag.} (N-1,-1, \cdots,-1 )\right]\non
&=&
\left(
\begin{array}{c|c}
 (q_{\rm R})_{11} & e^{-i\varphi/2} (q_{\rm R})_{1j} \\ \hline
e^{+i\varphi/2} (q_{\rm R})_{i1} & (q_{\rm R})_{ij}
\end{array}
\right) 
\eeq
with $i,j = 2, \cdots , N$.
After complete encirclement ($\varphi=2\pi$), they become
\beq
q_{\rm L} &\to& 
\left(
\begin{array}{c|c}
 e^{- ((2N-3)/N) \pi i}  (q_{\rm L})_{11}  & e^{-i ((N-3)/N) \pi i}  (q_{\rm L})_{1j}  \\ \hline
 e^{ -i((N-3)/N) \pi i}  (q_{\rm L})_{i1}  & e^{ 3 \pi i/N}  (q_{\rm L})_{ij} 
\end{array}
\right)
= 
\left(
\begin{array}{c|c}
 -  (q_{\rm L})_{11}  &  (q_{\rm L})_{1j}  \\ \hline
     (q_{\rm L})_{i1}  & - (q_{\rm L})_{ij} 
\end{array}
\right) \mbox{ (for $N$=3) } 
,\non
q_{\rm R} 
&\to&
\left(
\begin{array}{c|c}
    (q_{\rm R})_{11} & -  (q_{\rm R})_{1j} \\ \hline
 - (q_{\rm R})_{i1} &  (q_{\rm R})_{ij}
\end{array}
\right) .  \label{eq:gAB-light}
\eeq

For the case of $N=3$,
even numbers of encirclements 
give a trivial action.

%%%%%%%%%%%%
\section{Energetics of vortices}\label{sec:energy}

In this section, we calculate the leading contributions to 
tensions of vortices, in particular of 
chiral non-Abelian vortices $(1,0)$ and $(0,\pm 1)$, 
non-Abelian semi-superfluid vortices $(1,1)$,
and non-Abelian axial vortices $(1,-1)$.
We also calculate 
the tension of a composite state of two chiral non-Abelian vortices 
$(1,0)$ and $(0,1)$ with different 
${\mathbb C}P^{N-1}$ orientations,  
to show that these orientations are energetically favored 
to be aligned to each other, while 
${\mathbb C}P^{N-1}$ orientations of 
 two chiral non-Abelian vortices 
$(1,0)$ and $(0,-1)$ are energetically favored 
to be orthogonal to each other.

%%%%%
\subsection{Two vortices with parallel ${\mathbb C}P^{N-1}$ orientations}

Since all vortices discussed in this model 
have global $U(1)_{\rm B}$ windings, 
the leading contributions
to their tensions 
are logarithmically divergent with 
coming from the kinetic term of $\Phi_{\rm R,L}$,
as usual for global vortices.

We consider the following asymptotic configuration characterized by a set of two integers
$\{k_{\rm L},k_{\rm R}\}$,
\be
\Phi_{\rm L} &\to& v\, {\rm diag}\,\left( e^{ik_{\rm L}\varphi},\,1,\,\cdots,1\right),\\
\Phi_{\rm R} &\to& v\, {\rm diag}\,\left( e^{ik_{\rm R}\varphi},\,1,\,\cdots,1\right),
\ee
as $r \to \infty$. 
We have taken the ${\mathbb C}P^{N-1}$ moduli 
of the $(1,0)$ vortex of the left chirality 
to be oriented to the first component without loss of generality, 
and  
we assume  
that of 
 the $(0,1)$ vortex of the right chirality to be aligned 
 to the $(1,0)$ vortex in this subsection. 
 The case that they are not aligned is discussed in the next subsection.
 
The gauge fields should be chosen in such a way that the kinetic energy 
of $\Phi_{\rm L,R}$ is minimized:
\be
A_i &\to& -\epsilon_{ij}\frac{(k_{\rm L}+k_{\rm R})x^j}{2Ng_sr^2}\,T_N.
\ee
Then, the scalar kinetic energy reads
\be
{\cal K} = {\rm Tr}\,\left[
{\cal D}_i\Phi_{\rm L}^\dag {\cal D}_i\Phi_{\rm L} + {\cal D}_i\Phi_{\rm R}^\dag {\cal D}_i\Phi_{\rm R}
\right] \to \frac{v^2}{r^2} \frac{N+1}{2N}F_N(k_{\rm L},k_{\rm R}),
\ee
where $F$ is given by
\be
F_N(k_{\rm L},k_{\rm R}) = k_{\rm L}^2 - \frac{2(N-1)}{N+1}k_{\rm L}k_{\rm R} + k_{\rm R}^2
= (k_{\rm L}-k_{\rm R})^2 + \frac{4}{N+1}k_{\rm L}k_{\rm R}.
\ee
We thus find that 
the leading term of the tension is given by
\be
K = 2\pi \int^\Lambda dr\, r{\cal K} = \frac{(N+1)\pi v^2}{N} F_N(k_{\rm L},k_{\rm R}) \log \Lambda,
\ee
where $\Lambda$ is an IR cutoff parameter, or 
the size of the system. 

We have $F_N(1,0) = F_N(0,1) = 1$ for a single chiral non-Abelian vortex $(1,0)$ or $(0,1)$, 
and 
$F_N(1,1) = 4/(N+1)$ for 
a single non-Abelian semi-superfluid vortex $(1,1)$.
Comparing these two, we find $F_N(1,0) < F_N(1,1)$ for $N \le 2$, $F_N(1,0) = F_N(1,1)$ for $N=3$ (relevant to QCD), and
$F_N(1,0) > F_N(1,1)$ for $N\ge 4$.

By using these results, 
we can discuss whether 
 two separated chiral non-Abelian vortices 
with the opposite chiralities 
$(1,0)$ and $(0,1)$
are energetically bound to 
a single non-Abelian semi-superfluid vortex
$(1,1)$ or not.
To this end,
we note that 
when the vortices
$(1,0)$ and $(0,1)$ are infinitely separated, 
the tension of such a composite state is proportional to 
a sum of the tensions of individual vortices: 
$F_N(1,0) + F_N(0,1) =2$.
Comparing this with 
$F_N(1,1)$ of 
a single non-Abelian semi-superfluid vortex,
we find 
\beq
  F_N(1,0) + F_N(0,1) (=2) 
  \left\{\begin{array}{c} 
  = F_N(1,1)  = 2  \hspace{1.5cm} \mbox{ for } N=1 \\
  > F_N(1,1) = \displaystyle{\frac{4}{N+1}} 
  \hspace{0.7cm} \mbox{ for } N \geq 2
  \end{array}\right. .\label{eq:interactions}
\eeq
This implies that two chiral non-Abelian vortices 
with the opposite chiralities
$(1,0)$ and $(0,1)$ attract each other 
for $N \geq 2$, 
while there is no force between them at this order for $N=1$.
The latter  corresponds to two-component BECs 
in which the absence of the leading order interaction is in fact known
\cite{Eto:2011wp}.

In a similar way, we can discuss
the stability of a non-Abelian axial vortex $(1,-1)$.
When the vortices
$(1,0)$ and $(0,-1)$ are infinitely separated, 
the tension of such a composite state is proportional to
$F_N(1,0) + F_N(0,-1) =2$.
Comparing this with 
$F_N(1,-1)=4N/(N+1)$ of 
a single non-Abelian axial vortex,
we find 
\beq
  F_N(1,0) + F_N(0,-1) (=2) 
  \left\{\begin{array}{c} 
  = F_N(1,-1)  = 2  \hspace{1.5cm} \mbox{ for } N=1 \\
  < F_N(1,-1) = \displaystyle{\frac{4N}{N+1}} 
   \hspace{0.7cm} \mbox{ for } N \geq 2
  \end{array}\right. ,\label{eq:interactions}
\eeq
implying that two chiral non-Abelian vortices 
with the opposite chiralities
$(1,0)$ and $(0,-1)$ repel each other 
for $N \geq 2$, 
while there is no force between them at this order for $N=1$.
We thus have found that for $N \geq 2$  
the non-Abelian axial vortex $(1,-1)$ is unstable to decay
into $(1,0)$ and $(0,-1)$ chiral non-Abelian vortices.

\subsection{Two vortices with 
orthogonal ${\mathbb C}P^{N-1}$ orientations}

In this subsection, 
we take the ${\mathbb C}P^{N-1}$ moduli 
of the $(1,0)$ and $(0,1)$ vortices 
to be orthogonal to each other, 
which is possible for $N\geq 2$.
To this end, without loss of generality, we consider the following asymptotic configuration characterized by the set of two integers
$\{k_{\rm L},k_{\rm R}\}$,
\be
\Phi_{\rm L} &\to& v\, {\rm diag}\,\left( e^{ik_{\rm L}\varphi},\,1,\,\cdots,1\right),\\
\Phi_{\rm R} &\to& v\, {\rm diag}\,\left( 1,\,e^{ik_{\rm R}\varphi},\,\cdots,1\right),
\ee
as $r \to \infty$.
In this case, 
the gauge fields should be chosen in such a way that the kinetic energy 
of $\Phi_{\rm L,R}$ is minimized:
\be
A_i &\to& -\epsilon_{ij}\frac{k_{\rm L}x^j}{2Ng_sr^2}\,T_N-\epsilon_{ij}\frac{k_{\rm R}x^j}{2Ng_sr^2}\,T_N',
\ee
with $T_N' = {\rm diag}(-1,N-1,-1,\cdots,-1)$.
Then, the scalar kinetic energy reads
\be
{\cal K} \to \frac{v^2}{r^2} \frac{N+1}{2N}G_N(k_{\rm L},k_{\rm R}),
\ee
with $G$ defined by
\be
G_N(k_{\rm L},k_{\rm R}) = k_{\rm L}^2 + \frac{1}{N+1}k_{\rm L}k_{\rm R} + k_{\rm R}^2 .
\ee
Thus, the leading contribution to the tension of the composite state 
can be calculated, to give
\be
K = 2\pi \int^\Lambda dr\, r{\cal K} = \frac{(N+1)\pi v^2}{N} G_N(k_{\rm L},k_{\rm R}) \log \Lambda.
\ee

The tension of a set of two non-Abelian chiral vortices $(1,0)$ and $(0,1)$ with the ${\mathbb C}P^{N-1}$ orientations orthogonal to each other 
is thus found to be proportional to $G_N(1,1) = 2 + 1/(N+1)$. Since we have an inequality
\beq 
F_N(1,1) =\displaystyle{\frac{4}{N+1}}  < G_N(1,1) = 2 + \displaystyle{\frac{1}{N+1}}
\eeq
for all $N (\geq 2)$, the chiral non-Abelian 
vortices with aligned ${\mathbb C}P^{N-1}$ orientations are energetically more favored than those with
orthogonal orientations, 
implying that 
their ${\mathbb C}P^{N-1}$ moduli attract each other, 
to be aligned.

Again, in a similar way, we can discuss
the case of two chiral non-Abelian vortices with the opposite
 chiralities
 $(1,0)$ and $(0,-1)$.
 In this case,  we have an inequality 
\beq
 F_N(1,-1)  =\displaystyle{\frac{4N}{N+1}}  > G_N(1,-1) =2 - \displaystyle{\frac{1}{N+1}}
\eeq 
for all $N(\geq 2)$. 
Thus, two chiral non-Abelian vortices $(1,0)$ and $(0,-1)$ with orthogonal ${\mathbb C}P^{N-1}$ orientations are energetically more favored than those with
aligned orientations, 
implying that 
their ${\mathbb C}P^{N-1}$ moduli
 repel each other.
 If we separate them infinitely, the tension 
 becomes $G_N(1,0) + G_N(0,-1) = 2$. 
The inequality
\beq 
 G_N(1,0) + G_N(0,-1) = 2 > G_N(1,-1) =2 - \displaystyle{\frac{1}{N+1}}
\eeq
implies that 
the two chiral non-Abelian vortices $(1,0)$ and $(0,-1)$ with orthogonal ${\mathbb C}P^{N-1}$ orientations 
attract each other, forming a bound state.
It is, however, a highly nontrivial dynamical question 
remaining as a future problem. 

%
%%%%
%%%%%%%%%%%%
\section{Vortex-domain wall composites}\label{sec:chiral-sym-br}

We consider the case of 
$\gamma_1, \gamma_2, \gamma_3 \neq 0$ in which 
the axial and chiral symmetries are explicitly broken. 
This breaking induces domain walls attached to 
the vortices.

%%%%
\subsection{Chiral non-Abelian vortices attached by chiral domain walls}\label{sec:confinement}
\begin{figure}[t]
\begin{center}
\includegraphics[width=15cm]{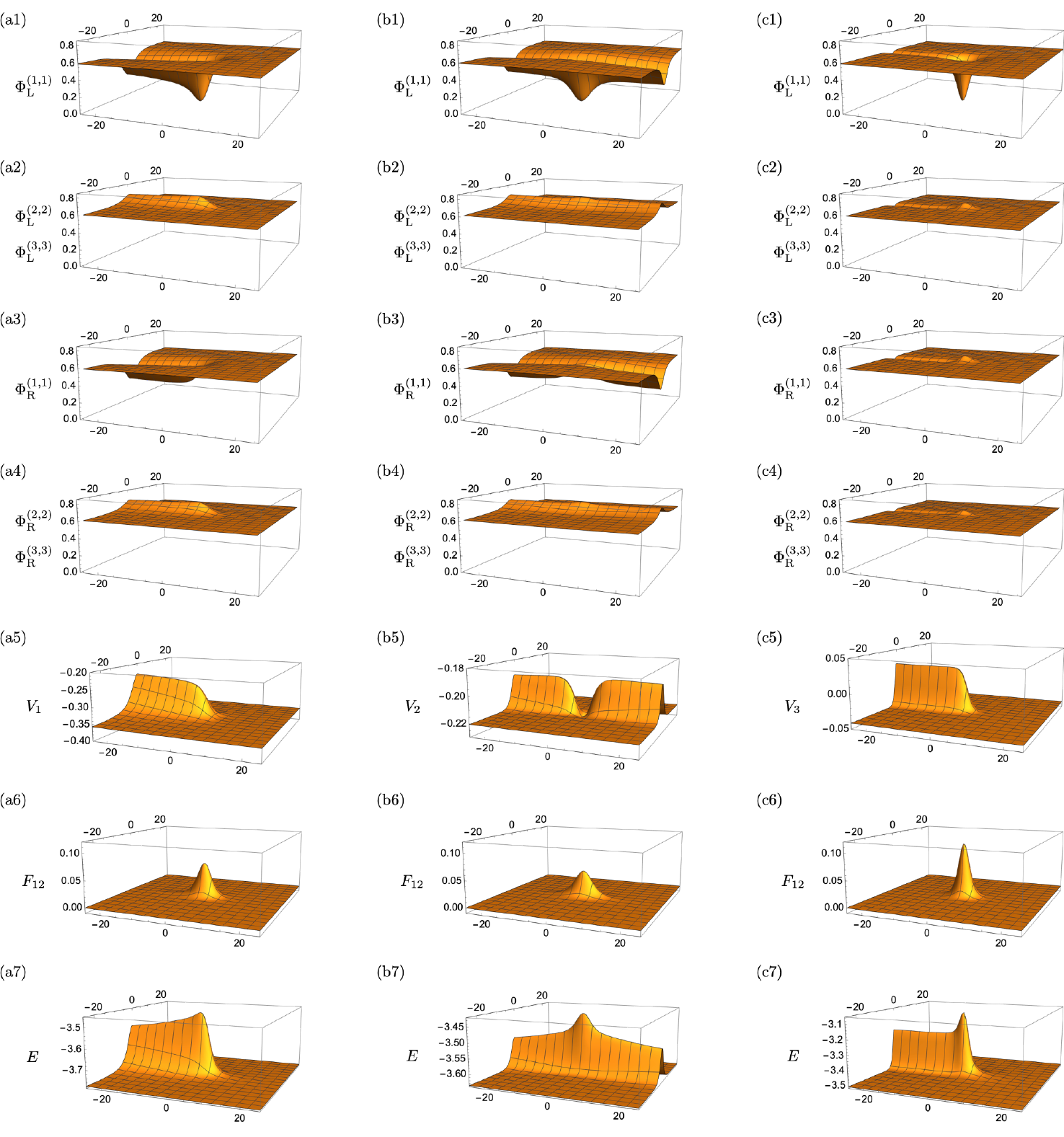}
\caption{
The profile functions $|\Phi_{\rm L,R}^{(i,i)}|^2$ $(i=1,2,3)$
of the vortex-wall composites. The left-most, middle, and right-most
columns have $(\gamma_1,\gamma_2,\gamma_3) = (-0.1,0,0)$, $(0,-0.1,0)$, $(0,0,-0.2)$, respectively.
The other parameters are common for all cases as $(m,\lambda_1,\lambda_2,\lambda_3,\lambda_4,g) = (\sqrt2,1,1,1,0,1)$.
}
\label{fig:vortex_wall}
\end{center}
\end{figure}

Let us turn on 
$\gamma_1, \gamma_2, \gamma_3 \neq 0$ 
to see their effects on vortices. 
In this subsection, 
we consider a 
chiral non-Abelian vortex.
In Fig.~\ref{fig:vortex_wall}, 
we present numerical simulations 
in the case that either of 
$\gamma_1, \gamma_2, \gamma_3$ 
is nonzero. 
We can clearly see that 
one chiral non-Abelian vortex 
is attached by one or two domain walls.
The three columns correspond 
 from the left to the right 
to
$(\gamma_1, \gamma_2, \gamma_3)=
(*,0,0), (0,*,0), (0,0,*)$, respectively.
In the middle column $\gamma_2 \neq 0$, 
the vortex is attached by 
the two domain walls  with the same tension, 
and thus the configuration is stable.
In left-most and right most cases, 
the vortex is attached by 
one domain wall from one side.
The wall pulls the vortex and the configuration is unstable, 
but it is static in the comoving frame.

To see why this happens, 
we consider an infinitely large circle 
parametrized by the spatial angle $\varphi$ 
encircling vortices. 
Let us substitute the chiral non-Abelian vortex ansatz
of either the left chirality in Eq.~(\ref{eq:ansatz})
or the right chirality in Eq.~(\ref{eq:ansatz2}) 
to the potential term in Eq.~(\ref{eq:GL}), 
with replacing the spatial angle $\varphi$ 
by a function $\phi(\varphi)$ 
depending on the angle $\varphi$ 
with the boundary condition
$\phi(\varphi=0) =0$ 
and $\phi(\varphi=2\pi) = 2\pi$.   
It can be evaluated at spatial infinities as
\beq
 V = 2( \gamma_1 +  \gamma_3)\cos \phi(\varphi) 
 + 2 \gamma_2 \cos (2\phi(\varphi)) .
\eeq
Together with the gradient term, 
the effective energy for $\phi$ on the large circle 
at infinity encircling the chiral non-Abelian axial vortex 
 becomes  
${\cal E}_{\rm eff} = v^2 (\partial_{\varphi} \phi)^2 + V$. 
This is the double sine-Gordon model.
Note that the periodicity is $2\pi$ in contrast to 
the case of non-Abelian axial vortices in 
Eq.~(\ref{eq:DSG-half}) in which the periodicity was $\pi$.

First, let us consider the absence of $\gamma_3$ ($\gamma_3=0$).
The case of 
 $(\gamma_1, \gamma_2 , \gamma_3  ) = 
 (\gamma_1,0,0)$ corresponds to 
 the sine-Gordon model, 
 the case of 
$(\gamma_1, \gamma_2 , \gamma_3  ) = 
 (0, \gamma_2 , 0 )$ 
 does to the sine-Gordon model with a half periodicity,
and the case of 
$(\gamma_1, \gamma_2 , \gamma_3  ) 
=(\gamma_1, \gamma_2 , 0 ) $  
 to the double sine-Gordon model. 
In the first case, a single chiral non-Abelian vortex 
is attached by a single sine-Gordon soliton, 
and thus is confined as shown in Fig.~\ref{fig:vortex_wall}(a1)--(a7).
In the second case, 
it is attached by two sine-Gordon solitons (or domain walls) 
of the same tension 
from the opposite sides, and therefore 
the composite state is stable, see Fig.~\ref{fig:vortex_wall}(b1)--(b7).
\label{footnote:p-wave}
This case is a non-Abelian generalization of 
chiral P-wave superconductors, 
for which the GL theory is described by 
a $U(1)$ gauge theory coupled with 
two complex scalar fields $\Phi_1$ and $\Phi_2$ 
with a potential term $V \sim 
(\Phi_1^*)^2 (\Phi_2)^2 +{\rm c.c.}$ 
\cite{PhysRevLett.80.5184,PhysRevB.86.060514}. 
In the third case, it
is attached by two sine-Gordon solitons (or domain walls),
but how they attach depends on the parameters 
$\gamma_1, \gamma_2$
as classified in Refs.~\cite{Eto:2018hhg,Eto:2018tnk} 
in the context of two Higgs doublet models.
In these cases, 
the domain walls attached to the chiral non-Abelian vortex  
are non-Abelian sine-Gordon solitons 
carrying 
${\mathbb C}P^{N-1}$ moduli \cite{Nitta:2014rxa,Eto:2015uqa}.
This can be clearly seen in Fig.~\ref{fig:vortex_wall};
The component $\Phi_{\rm L}^{11}$ has the vortex winding 
[(a1) and (b1)].
If one looks at the right condensation 
along the domain wall,  
one finds that 
 $\Phi_{\rm R}^{11}$ is concave [Fig.~\ref{fig:vortex_wall} (a3) and (b3)] while
 $\Phi_{\rm R}^{22}$ and  
 $\Phi_{\rm R}^{33}$ are convex [Fig.~\ref{fig:vortex_wall}(a4) and (b4)].
 The same happens in the left condensations 
 along the wall far apart from the vortex 
 [Fig.~\ref{fig:vortex_wall}(a1), (a2), (b1) and (b2)]. 
 Thus, the $SU(3)$ symmetry is spontaneously broken down 
 to $SU(2) \times U(1)$ along the domain wall, 
 resulting in the ${\mathbb C}P^2$ NG modes 
 localized on the wall  
 or attributing the ${\mathbb C}P^2$ moduli.
These moduli match those 
in Eq.~(\ref{eq:CPN-1})
of the vortex along the junction line of the vortex and domain walls.

When only $\gamma_3$ is present, 
$(\gamma_1, \gamma_2 , \gamma_3  ) = 
 (0, 0 ,  \gamma_3   )$,
 one sine-Gordon soliton is attached to one chiral non-Abelian vortex
 as shown in Fig.~\ref{fig:vortex_wall}(c1)--(c7).
 This is Abelian, carrying no moduli. 
 Indeed, the profile functions 
 $|\Phi_{\rm L,R}^{(i,i)}|^2$ behave almost the same along the domain wall far from
 the vortex core. This implies that $\Phi_{\rm L,R}$ are proportional to the identity, and so
 no symmetries are broken by the domain wall.

  When all $\gamma_{1,2,3}$ are present, 
  there appears 
either attraction or repulsion among the domain walls 
attached to the vortex, depending on its sign. 
If it is attraction, the domain walls form a composite domain wall 
\cite{Eto:2013hoa,Eto:2013bxa}, 
thus confining the chiral non-Abelian vortex.
If repulsion, the chiral non-Abelian vortex 
is attached by two domain walls 
with different tensions from opposite sides.
Such details of the domain wall structure 
are worth to study on their own, 
but are not relevant in the following 
subsections for vortex molecules, 
as explained below. 

If we do the same for the usual non-Abelian semi-superfluid vortex in
 Eq.~(\ref{eq:ansatz0}), 
 there is no potential term, implying that no domain wall 
 is attached to the usual non-Abelian semi-superfluid vortex.

%%%%%%
\subsection{Decay of Abelian and non-Abelian axial vortices}\label{sec:decay}
Here we discuss that Abelian and non-Abelian axial vortices 
are all unstable to decay into a set of chiral non-Abelian vortices 
once the axial and chiral symmetry breaking terms are turned on.

One non-Abelian axial vortex discussed in 
Sec.~\ref{sec:NAGV} is attached from the opposite sides 
by two (or four for $(\gamma_1,\gamma_2,\gamma_3)=(0,\gamma_2,0)$) domain walls extending to infinities,
 and thus decays 
into two chiral non-Abelian vortices 
each of which is 
attached by one (or two) chiral domain wall(s) as in Fig.~\ref{fig:NAVM}(a); one of the left chirality 
and the other of the right chirality with the opposite winding.
This decay process can be written as
\beq
 (1,-1) \to (1,0) + (0,-1).
\eeq
It is interesting to observe that 
there was no flux in the initial state while 
the final states contain fluxes. 
Similarly, the Abelian axial vortex is also unstable to decay as
\beq
 (N,-N) \to N (1,-1) \to N (1,0) + N(0,-1).
\eeq

Another example is  
a doubly-wound chiral non-Abelian vortex with 
the same chirality, say left. 
This is also attached by two (or four for $(\gamma_1,\gamma_2,\gamma_3)=(0,\gamma_2,0)$) 
chiral domain walls extending to infinities 
(but vortices of the same chirality are placed at the both L and R),  
and therefore it is also unstable against decay 
into two chiral non-Abelian vortices each of which is 
attached by one (or two) chiral domain wall(s) as in Fig.~\ref{fig:NAVM}(a).
This decay process can be written as
\beq
 (2,0) \to 2 (1,0).
\eeq
Similarly, an Abelian axial string having the minimum unit winding in 
$\Sigma$ can decay as
\beq
 (N,0) \to N (1,0).
\eeq
This decay was numerically simulated 
in the linear sigma model 
\cite{Eto:2013hoa,Eto:2013bxa}.   

%%%%%%%%%%%%
\section{Non-Abelian vortex molecules}\label{sec:chiralNAmolecule}

%%%%
\subsection{Structure of chiral non-Abelian vortex molecules}

Before discussing the effect of explicit breaking terms $\gamma_{1,2,3} \neq 0$ for general case, 
let us make a comment on the interaction between 
chiral non-Abelian vortices 
for  $\gamma_{1,2,3} = 0$  
in the case of $N=1$, in which the system reduces to 
two-component BECs.
In this case, 
the interaction energy between 
$(1,0)$ and $(\pm 1,0)$ vortices at distance $R$ 
 is well known $E_{\rm int}\sim \pm \log R$. 
 Thus, a vortex and (anti-)vortex repel (attract) each other 
 as usual for single component global (superfluid) vortices. 
 On the other hand, 
the interaction energy between 
$(1,0)$ and $(0,1)$ vortices at distance $R$ 
vanishes at the leading order, 
to be consistent with Eq.~(\ref{eq:interactions}),  
and the next leading order 
is $E_{\rm int}\sim \lambda \log R/R^2$
\cite{Eto:2011wp} 
with $\lambda$ being 
$\lambda_3$ and/or $\lambda_4$ in Eq.~(\ref{eq:GL}) 
(reducing the same term for $N=1$).
Thus, it can be either repulsive ($\lambda >0$) 
or attractive ($\lambda <0$).
Once we introduce the explicit breaking terms 
$\gamma_{1,2,3} \neq 0$,   
a pair of $(1,0)$ and $(0,1)$ vortices forms a molecule 
in which constituents are separated at finite distance,  
when they are repulsive ($\lambda >0$) 
\cite{Kasamatsu:2004tvg,Eto:2017rfr}.
They collapse to form an Abelian vortex $(1,1)$
when they are attractive ($\lambda <0$).

\begin{figure}
    \centering
    \begin{tabular}{cc}
    \includegraphics[height=3cm]{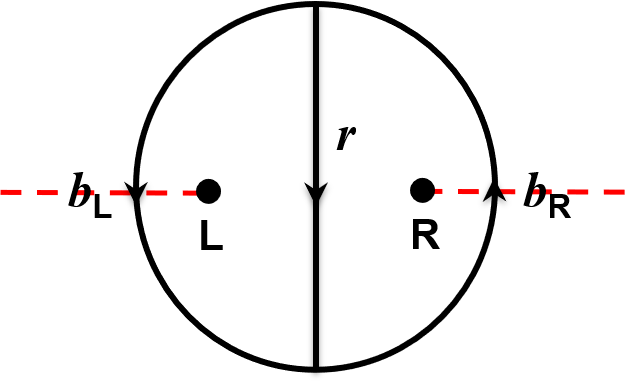}
    &     \includegraphics[height=3cm]{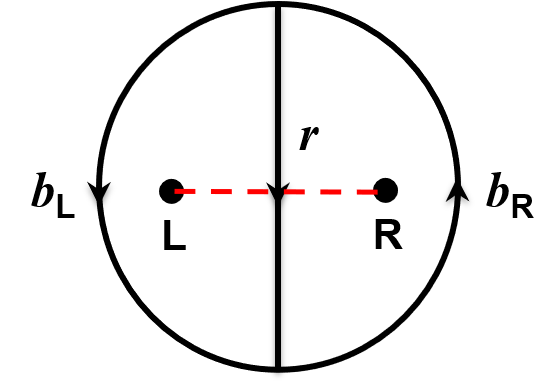}\\
    (a) & (b)
    \end{tabular}
    \caption{Pairs of chiral non-Abelian vortices
       (a) attached by domain walls extending to infinities, 
    leading to the instability against a decay, and  
    (b) forming a chiral non-Abelian vortex molecule connected by a domain wall. 
    For the both cases, 
    left (right) chiral non-Abelian vortices placed at L and R are  
    encircled by the closed loops $b_{\rm L} -r$ and $b_{\rm R} +r$, respectively.
    (a) They have the opposite windings $(1,0)$ and $(0,-1)$, and are attached by domain walls extending to infinities, leading the instability against decay. 
    (b) These vortices have the same windings $(1,0)$ and $(0,1)$, and 
    are connected by a domain wall denoted by a red broken line 
    to form a vortex molecule.
    }
    \label{fig:NAVM}
\end{figure}

Here, we show that the chiral non-Abelian vortices 
$(1,0)$ and $(0,1)$ can form a molecule.
They have the same color magnetic fluxes.
Now we put a $(1,0)$-vortex on the left at ``L''
and a $(0,1)$-vortex on the right at ``R'' 
in Fig.~\ref{fig:NAVM} (b). We assume that 
the ${\mathbb C}P^{N-1}$ orientations of these vortices are the same.
The loop $b_{\rm R} + r$ encircles the $(0,1)$-vortex while 
the one $b_{\rm L}-r$ encircles the $(1,0)$-vortex.
The large loop $b_{\rm R}+b_{\rm L}$ encircles the both of them.
% ---   figure   ---%
%---   figure   ---%

Along each of the large half circles $b_{\rm L}$ and $b_{\rm R}$, 
the vector transformations, {\it i.~e.~},
the color gauge transformation
and $U(1)_{\rm B}$ transformation act as
\beq
b_{\rm L}, b_{\rm R}: \quad
 && g_C(\varphi) = e^{{i \over 2N} F(\varphi \mp {\pi \over 2}) T_N}  , \quad
 F(0)=0, \quad F(\pi)= 2\pi\\
&&  e^{i\theta_{\rm B}(\varphi)} = e^{i B(\varphi\mp {\pi \over 2}) } ,\quad 
 B(0) = 0 ,\quad B(\pi)= \pi/N,
\eeq
respectively, 
where $F$ and $B$ are  monotonically increasing 
functions (linear functions).
On the other hand, along the path $r$, we have
\beq
r: \quad
 && U_{\rm L}^\dagger = U_{\rm R} = e^{{i \over 2N} R(y) T_N}  , \quad
 R(-\infty)=0, \quad R(+\infty)= 2\pi\\
&&  e^{i\theta_{\rm A}(y)} = e^{iA(y)},\quad 
 A(-\infty) = 0 , \quad A(+\infty)= \pi/N,
\eeq
respectively, 
where we have parametrized the path $r$ by the coordinate $y$, 
and $R$ and $A$ are monotonically increasing functions.

Therefore, along the path $r$, 
there appears 
a (composite) domain wall stretching between
the $(1,0)$- and $(0,1)$-vortices 
once $\gamma_{1,2,3}$ are turned on.
The internal structure of the domain wall 
depends on the values of $\gamma_{1,2,3}$, 
as discussed in Sec.~\ref{sec:confinement}.
In the presence of only $\gamma_{1,3}$, 
there exists one domain wall between the vortices, 
while 
there are two domain walls in the presence of 
 $\gamma_2$. 
 Nevertheless, all domain walls must be stretched between 
 the two vortices since there is no wall 
 along the path $b_1+b_2$ encircling the whole configuration.  
Dynamically, the domain wall tension pulls 
these chiral non-Abelian vortices and combine them 
to a single non-Abelian semi-superfluid vortex.
Thus, these chiral non-Abelian vortices are confined 
to a ``mesonic'' configuration, which is nothing but 
a non-Abelian semi-superfluid vortex.
We can express this confining process by 
\beq
 (1,0) + (0,1) \to (1,1).
\eeq

\begin{figure}[th]
\begin{center}
\includegraphics[width=15cm]{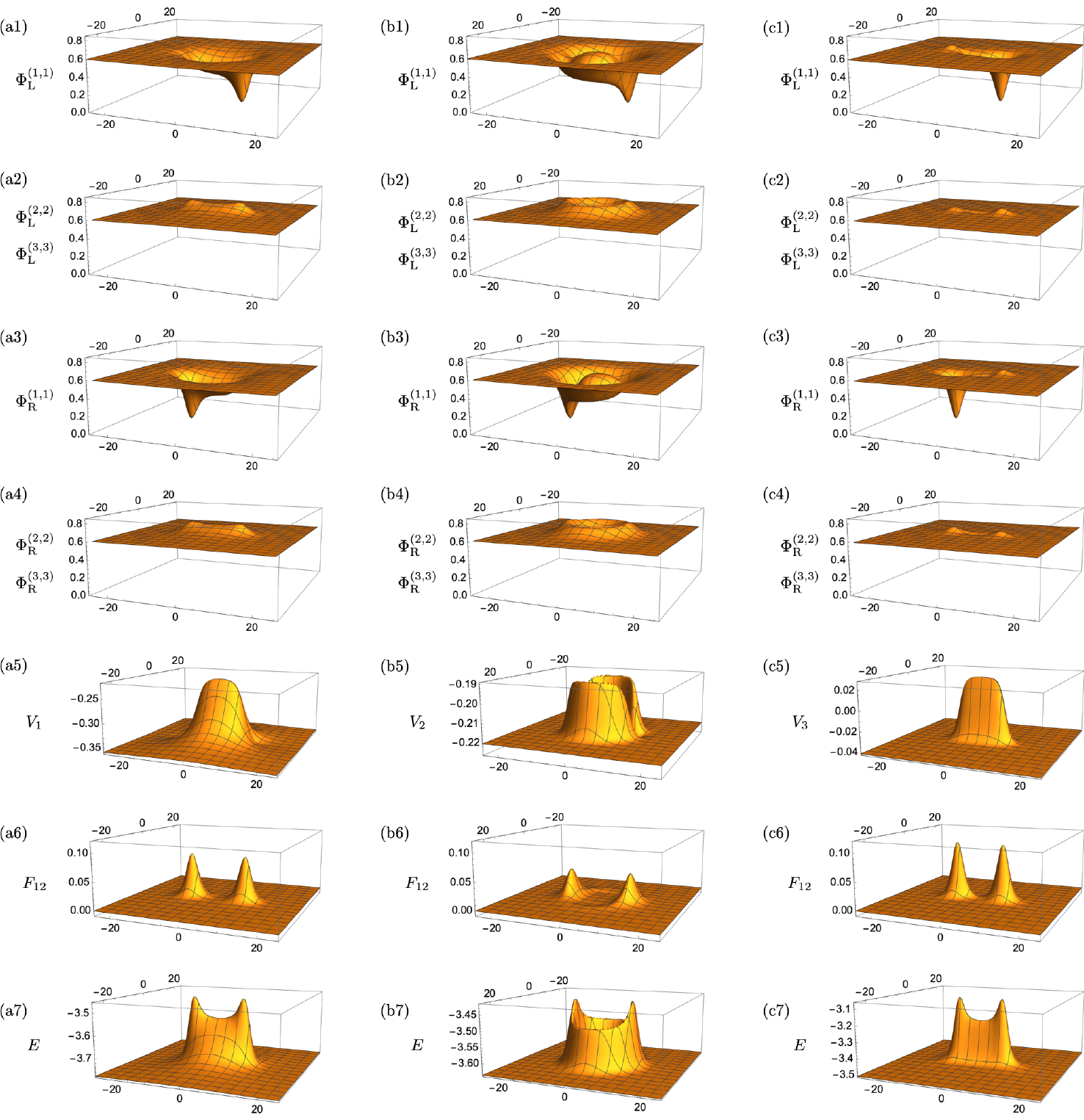}
\caption{
The profile functions $|\Phi_{\rm L,R}^{(i,i)}|^2$ $(i=1,2,3)$
of a pair of the $(1,0)$ and $(0,1)$ chiral non-Abelian vortices.
They are stretched by a chiral domain wall(s).
The parameter choice of the left-most, middle, and right-most
columns are $(\gamma_1,\gamma_2,\gamma_3) = (0.01,0,0)$, $(0,0.01,0)$, $(0,0,0.2)$, respectively.
The other parameters are common for all cases as $(m,\lambda_1,\lambda_2,\lambda_3,\lambda_4,g) = (\sqrt2,1,1,1,0,1)$.
}
\label{fig:vmolecule_sol}
\end{center}
\end{figure}
In Fig.~\ref{fig:vmolecule_sol}, we present numerical simulations 
of a pair of chiral non-Abelian vortices $(1,0)$ and $(0,1)$ 
separated at finite distance. 
Clearly one can see that the domain walls are stretched between 
them. 
In the case of 
$(\gamma_1,\gamma_2,\gamma_3) = (*,0,0)$ (the left column),
a single non-Abelian sine-Gordon kink is stretched between them.
In the case of 
$(\gamma_1,\gamma_2,\gamma_3) = (0,*,0)$ (the middle column), 
two domain walls are stretched between them forming a ring, 
like vortex molecules in a chiral P-wave superconductor, 
see footnote \ref{footnote:p-wave}.
Finally, in the case of 
$(\gamma_1,\gamma_2,\gamma_3) = (0,0,*)$ (the right column), 
a single Abelian sine-Gordon kink is stretched between them. 
%%%%

The two vortices are linearly confined and 
these configurations are on a way to collapse. 
It is an open question  
whether,
as the case of two-component BECs ($N=1$), 
these constituents can be separated at finite distance 
and an internal structure of 
the molecule is visible 
in certain parameter region.
Possibly, it may occur   
when $\lambda_4(>0)$ is large enough 
for which the $(1,0)$ and $(0,1)$ vortices would repel each other 
at short distances.
Explicitly constructing numerical solutions remains a future problem.

%%%%%%%%%%%%%
\subsection{Generalized Aharonov-Bohm phases}
Let us discuss generalized AB phases around a vortex 
molecule  in Fig.~\ref{fig:NAVM}(b).  
We restrict to $N=3$ relevant for the CFL phase.

When the light quarks encircle the (1,0)-vortex at L along the path $b_{\rm L} -r$, they receive the generalized 
AB phases 
\beq
q_{\rm L} \to 
\left(
\begin{array}{c|c}
 -  (q_{\rm L})_{11}  &  (q_{\rm L})_{1j}  \\ \hline
     (q_{\rm L})_{i1}  & - (q_{\rm L})_{ij} 
\end{array}
\right) 
,\quad
q_{\rm R} 
\to
\left(
\begin{array}{c|c}
    (q_{\rm R})_{11} & -  (q_{\rm R})_{1j} \\ \hline
 - (q_{\rm R})_{i1} &  (q_{\rm R})_{ij}
\end{array}
\right) .  \label{eq:AB-L}
\eeq
These constitute a ${\mathbb Z}_2$ group, 
which is a color non-singlet.
On the other hand, when they encircle the (0,1)-vortex at R  
along the path $b_{\rm R} +r$, they receive the generalized 
AB phases 
\beq
q_{\rm L} \to 
\left(
\begin{array}{c|c}
   (q_{\rm L})_{11}  & - (q_{\rm L})_{1j}  \\ \hline
 -    (q_{\rm L})_{i1}  &  (q_{\rm L})_{ij} 
\end{array}
\right) 
,\quad
q_{\rm R} 
\to
\left(
\begin{array}{c|c}
  -  (q_{\rm R})_{11} &   (q_{\rm R})_{1j} \\ \hline
  (q_{\rm R})_{i1} & - (q_{\rm R})_{ij}
\end{array}
\right) ,  \label{eq:AB-R}
\eeq
constituting a ${\mathbb Z}_2$ group, 
which is a color non-singlet.

Thus, when they encircle the both vortices along the large circle 
$b_{\rm L}+b_{\rm R}$, 
they receive the generalized AB phases 
\beq
 q_{\rm L} \to -q_{\rm L}, \quad q_{\rm R} \to -q_{\rm R}, \label{eq:AB-LR}
\eeq
constituting a ${\mathbb Z}_2$ group.
This is a color singlet.
These generalized AB phases are precisely 
those of a single non-Abelian semi-superfluid vortex 
\cite{Yasui:2010yw,Fujiwara:2011za}.
Interestingly, 
the light quarks can detect the color of fluxes of 
the chiral non-Abelian vortices $(1,0)$ and $(0,1)$ at the large distance 
by the generalized AB phases in  
Eqs.~(\ref{eq:AB-L}) and (\ref{eq:AB-R}) 
which are color non-singlets, 
but they 
cannot do that of the whole molecule by the generalized AB phase in Eq.~(\ref{eq:AB-LR}) 
which is a color singlet.

As for heavy quarks, they detect only gauge fields.
Thus, they do not distinguish  the (1,0)- and (0,1)-vortices 
unlike the light quarks, 
since the gauge structures are identical between 
these two vortices 
carrying exactly the same color magnetic fluxes.
Therefore, when they encircle either of  the (1,0)- and (0,1)-vortices, 
they receive the AB phases in Eq.~(\ref{eq:AB-phase-Q}) 
which is a color non-singlet,
while 
when they encircle the both of them along the path $b_{\rm L}+b_{\rm R}$, 
they receive the AB phases in Eq.~(\ref{eq:AB-phase-Q-twice}) 
which is a color singlet.
The latter forming  
a color singlet ${\mathbb Z}_3$ group 
are precisely 
those of a single non-Abelian vortex 
\cite{Chatterjee:2018nxe}.

In summary, 
chiral non-Abelian vortices 
are not confined and can exist alone 
when $\gamma_{1,2,3}=0$, 
while 
they are confined 
when $\gamma_{1,2,3}\neq 0$.
In the deconfined phase, 
chiral non-Abelian vortices 
exhibit color non-singlet (generalized) AB phases 
so that the light/heavy 
quarks can detect the colors of magnetic fluxes of these vortices 
at large distances. 
In the confined phase, 
chiral non-Abelian vortices 
exhibit only color singlet (generalized) AB phases 
so that the light/heavy 
quarks cannot detect the colors of magnetic fluxes of these vortices 
at large distances. 
Thus, stable states exhibit color-singlet (generalized) AB phases.

The opposite is not always true.
Not all states with color-singlet (generalized) AB phases 
can exist stably in the confined phase.
For instance, the two examples 
 $(1,-1)$ and $(2,0)$ 
in Sec.~\ref{sec:decay} 
exhibit 
color-singlet (generalized) AB phases; 
the $(2,0)$ made of two $(1,0)$ with the same color magnetic 
fluxes
exhibits the singlet AB phase for heavy quarks 
in Eq.~(\ref{eq:AB-phase-Q-twice}) 
and the trivial phases for light quarks 
obtained from two successive phases of Eq.~(\ref{eq:gAB-light});
$(Q_{\rm L}, Q_{\rm R}, q_{\rm L}, q_{\rm R}) 
 \to (\epsilon^2 Q_{\rm L}, \epsilon^2 Q_{\rm R}, q_{\rm L}, q_{\rm R})$, which are color singlet. 
Nevertheless,  
they are attached by the two chiral domain walls extending to infinities 
as in Fig.~\ref{fig:NAVM}(a),  
and are unstable against decay 
into two chiral non-Abelian vortices each of which is 
attached by one (or two) chiral domain wall(s): 
 $(1,-1) \to (1,0) + (0,-1)$ and
 $(2,0) \to 2 (1,0)$. 
 The $(N,0)$ vortex made of $N$ $(1,0)$ vortices 
 with all different color magnetic fluxes give generalized 
 AB phases as $(Q_{\rm L}, Q_{\rm R}, q_{\rm L}, q_{\rm R}) 
 \to (Q_{\rm L}, Q_{\rm R}, -q_{\rm L}, q_{\rm R})$, 
 which are color singlet. 
 It is, however, broken as $(N,0) \to N(1,0)$.

%%%%%%%%%%%%%%%%%%%%%%%%%%%%%%%%%%%
\section{Summary and discussion}\label{sec:summary}

In the CFL phase of dense QCD,
we have found  
 chiral non-Abelian vortices winding only around 
 either of left or right diquark condensation $\Phi_{\rm L}$ or  $\Phi_{\rm R}$ labeled by $(1,0)$ and $(0,1)$, respectively. 
 As can be expected from  $(1,0) = {1\over 2}[(1,1)+(1,-1)]$ 
 and  $(0,1) = {1\over 2}[(1,1)-(1,-1)]$,
  they carry half color magnetic fluxes 
  and half $U(1)_{\rm B}$ circulation 
of those of a non-Abelian semi-superfluid vortex labeled by $(1,1)$, 
and half  $U(1)_{\rm A}$ winding 
and half chiral circulation  
(around the sub-OPM ${\cal M}_{\rm A} = U(N)_{\rm L-R+A}$) 
of a non-Abelian axial vortex labeled by $(1,-1)$. 
A single chiral non-Abelian vortex 
carries ${\mathbb C}P^{N-1}$ orientational moduli in the internal space  corresponding to 
its color magnetic flux. 
We have discussed the energetics of vortices 
and have found that 
 ${\mathbb C}P^{N-1}$ orientations of  
two chiral non-Abelian vortices $(1,0)$ and $(0,1)$ 
are energetically aligned,
 while those with a chiral vortex $(1,0)$ and anti-vortex $(0,-1)$ 
  are energetically orthogonal to each other.
 Then, the two chiral non-Abelian vortices attract each other 
 forming bound states.
We have shown that 
chiral non-Abelian vortices exhibit the  
topological obstruction implying that 
the unbroken symmetry generators are not defined globally around the vortices,
and color non-singlet (generalized) AB phases 
implying that 
quarks at large distances 
can detect the colors of magnetic fluxes by encircling these vortices. 
In the presence of the axial and chiral symmetry breaking terms 
$\gamma_{1,2,3} \neq 0$, 
these vortices are confined by chiral domain walls,  
while 
they are deconfined in the absence of those terms. 
In the confined phase, 
two chiral non-Abelian vortices $(1,0)$ and $(0,1)$ with chiralities opposite to each other   
are connected by a chiral domain wall, constituting  
a mesonic bound state $(1,1)$ 
which is nothing but a non-Abelian semi-superfluid vortex, 
exhibiting only color singlet (generalized) AB phases 
implying that the quarks cannot detect the color of magnetic flux of such a bound state  
at large distances. 
We also have shown that the Abelian
axial vortices $(N,-N)$ and 
non-Abelian axial vortices $(1,-1)$ 
attached by chiral domain walls are both unstable to decay into 
a set of chiral non-Abelian vortices.

Before closing this paper, 
let us address several discussions and future directions.
\changed{ 
The confinement does not imply that the mesonic bound state neccesary 
collapses to a axisymmetric 
non-Abelian semi-superfluid vortex.
It remians as a futre problem to  numerically construct solutions of  vortex molecules in certain parameter regions, 
in which the constituent $(1,0)$ and $(0,1)$ vortices are separated
 at finite distances
At least 
the axial and chiral symmetry breaking terms should be 
relatively small.
At finite temperature, it does not have to be the case 
at least in 2+1 dimensions
because of the Berezinskii-Kosterlitz-Thouless (BKT) transition.
The BKT transition was explicitly shown 
in Ref.~\cite{Kobayashi:2018ezm} 
by numerical simulations  
for the Abelian case $N=1$. 
}

In this paper, we have constructed numerical solutions 
for single chiral non-Abelian vortices $(1,0)$ or $(0,1)$, 
in the absence of the axial and chiral symmetry breaking terms:
$\gamma_{1,2,3}=0$. 
In the presence of these terms, chiral domain walls are 
attached to them. 
\changed{
We have constructed solutions in the case that only one of
 $\gamma_{1,2,3}$ is nozero.
 }
In particular, if we turn on all $\gamma_{1,2,3}$'s, 
the situation is close to the two-Higgs doublet models 
\cite{Eto:2018hhg,Eto:2018tnk}. 
Explicitly constructing numerical solutions of 
such domain-wall vortex composites \changed{in general cases} 
remains as one of future problems.
A particularly important problem is to construct 
a vortex molecule $(1,0)$ + $(0,1)$. 
This would reduce to 
a single non-Abelian semi-superfluid vortex  
in the most parameter region because of the domain wall tension, 
but we should examine 
whether these two constituents can be separated 
in some parameter region particularly for small $\gamma_{1,2,3}$'s 
and/or small gauge coupling $g_s$ for which 
there is a repulsion between constituent vortices. 
This problem is important in a relation with 
higher-form symmetries 
discussed in the next paragraph.
Finally, we also should numerically verify   
decays of axial and chiral vortices 
as discussed in Sec.~\ref{sec:decay}, 
such as 
a non-Abelian axial vortex  $(1,-1) \to (1,0)+(0,-1)$ 
and   
an Abelian axial vortex $(N,-N) \to N(1,0)+N(0,1)$, 
as we did a similar problem in the linear sigma models 
\cite{Eto:2013hoa,Eto:2013bxa}.  
In particular, in the presence of the mass terms 
$\gamma_{1,2}\neq 0$, 
$2N$ domain walls attached to one Abelian axial vortex 
$(N,-N)$ constitute a composite wall 
as can be expected from Eq.~(\ref{eq:pot-abelian-axial}), 
and it is an open question whether this fact suppresses the decay.
       To perform simulations, we may do either 
       a relaxation method or real time dynamics. For the latter, we need a time-dependent GL theory.

Higher-form symmetries  
\cite{Gaiotto:2014kfa} related with 
a linking between Wilson loops and vortices 
are an indispensable tool 
to study phases of matter 
such as the so-called topological order.  
Higher form symmetries in the presence of 
non-Abelian semi-superfluid vortices 
and the absence or presence of a topological order 
of the CFL phase 
were studied 
in Refs.~\cite{Cherman:2018jir, Hirono:2018fjr, Hirono:2019oup,Hidaka:2019jtv,Cherman:2020hbe}. In this case, a linking between 
a Wilson loop and a non-Abelian semi-superfluid vortex is 
rather trivial in the sense that AB phases are color singlets.
 Contrary to this, 
a linking between 
a Wilson loop and a chiral non-Abelian vortex is 
non-trivial because AB phases are color non-singlets 
as we have seen in Sec.~\ref{sec:gAB}.
Thus, the phase separating a non-Abelian semi-superfluid vortex 
$(1,1)$ into two chiral non-Abelian vortices 
$(1,0)$ and $(0,1)$ may be characterized in terms of  
a higher-form symmetry.

As mentioned in introduction, 
in the context of  quark-hadron continuity, 
vortices penetrate through the CFL phase and 
hyperon nuclear matter~\cite{Alford:2018mqj, Chatterjee:2018nxe,
  Chatterjee:2019tbz, Cipriani:2012hr, Cherman:2018jir, Hirono:2018fjr, Hirono:2019oup, Cherman:2020hbe}. 
In particular, from the AB phases of quarks around vortices, 
one can conclude the existence of a boojum 
at which three hyperon vortices and three non-Abelian 
semi-superfluid vortices must meet 
 \cite{Chatterjee:2018nxe,Chatterjee:2019tbz, Cipriani:2012hr}.
This structure is modified if the deconfined phase 
is realized in the CFL phase.
In fact, this situation is similar to two-flavor quark matter 
\changed{(see Appendix \ref{sec:two-flavors}).}

Beyond the GL description, 
we could study fermion structure by 
the BdG formulation.
In fact, 
fermion zero modes were studied for 
non-Abelian semi-superfluid vortices in the BdG equation
  \cite{Yasui:2010yw,Fujiwara:2011za,Chatterjee:2016ykq},
  in which triplet Majorana fermion zero modes were found. 
  Such Majorana fermions endow these vortices a non-Abelian exchange statistics in $d=2+1$, turning them into non-Abelian anyons 
  \cite{Yasui:2010yh,Hirono:2012ad}. 
  Apparently, it is a very interesting question 
  whether fermion zero modes exist on 
  chiral non-Abelian vortices 
  and if so what is their exchange statistics.

The ${\mathbb C}P^2$ modes of the chiral non-Abelian vortex 
are probably non-normalizable, 
unlike those of 
a single non-Abelian semi-superfluid vortex 
 \cite{Eto:2013hoa,Eto:2009tr}.
However, around a constituent of a vortex molecule $(1,1)$, 
these modes may be normalizable because of a cut-off introduced by the presence of the other.
The ${\mathbb C}P^2$ modes are normalizable on the chiral domain wall  
\cite{Nitta:2014rxa,Eto:2015uqa} that connects the 
$(1,0)$ and $(0,1)$.
This fact together with a fact that 
the ${\mathbb C}P^2$ modes are normalizable
on a single non-Abelian semi-superfluid vortex 
 \cite{Eto:2013hoa,Eto:2009tr} may suggest 
that these modes are still normalizable 
around the vortex molecule $(1,1)$.

In this paper, we have turned off the electro-magnetic interaction 
and the strange quark mass.
Turning them on can be incorporated in the ${\mathbb C}P^2$ effective 
world-sheet Lagrangian of a single non-Abelian semi-superfluid vortex 
in Refs.~\cite{Vinci:2012mc} and \cite{Eto:2009tr}, 
respectively. 
This method may be applied to the case of 
a chiral non-Abelian vortex as well.

Finally, 
there are some interesting directions for
studying chiral domain walls.
One is a decay of chiral domain walls by 
quantum or thermal tunneling. 
In this case, a hole created on the domain wall world-volume 
is surrounded by an axial vortex (see  Sec.~10.5 of the review paper \cite{Eto:2013hoa}). For the minimum element of 
a chiral domain wall,
a hole should be surrounded by 
a chiral non-Abelian vortex studied in this paper.
The other direction is given by  
is the chiral non-Abelian
semi-superfluid vortices under magnetic field background which would be also
interesting in connection with the chiral anomaly.
The domain wall connecting $(1,0)$ and $(0,1)$ is made of the $\eta'$ meson
related to $U(1)_{\rm A}$, and $\eta'$ nontrivially changes along the direction perpendicular
to the $\eta'$ domain wall. Therefore, under the presence of magnetic field, the
domain wall should be magnetized as found in Refs.~\cite{Son:2007ny,Eto:2012qd}, see also Sec.~10.6 of Ref.~\cite{Eto:2013hoa}. 
Physical consequences of these domain walls are interesting to explore.

%%%%%%%%%%%%%%
\section*{Acknowledgments}

The work of M.~E.~ is supported in part 
by JSPS Grant-in-Aid for Scientific Research 
KAKENHI Grant No. JP19K03839, and
by MEXT KAKENHI Grant-in-Aid for 
Scientific Research on Innovative Areas
``Discrete Geometric Analysis for Materials Design'' No. JP17H06462 from the MEXT of Japan.
The work of M.~N.~is supported 
in part by JSPS KAKENHI Grant Number 18H01217.

\begin{appendix}

\section{Terminologies}\label{sec:term}

\changed{
In this Appendix, we summarize 
terminologies in this paper, which may be sometimes confusing.
}
\subsection{Abelian and Non-Abelian} 
 
 \changed{ 
The terminology ``Abelian vortices'' is used for 
vortices having winding in a $U(1)$ group.
In this paper, such a $U(1)$ group is 
either the baryonic symmetry $U(1)_{\rm B}$ or axial symmetry 
$U(1)_{\rm A}$.
A vortex winding around $U(1)_{\rm B}$ is called 
a Abelian superfluid vortex 
(Sec.~\ref{sec:Abelian-superfluid}), 
while one winding around 
$U(1)_{\rm A}$ is called an axial vortex 
(Sec.~\ref{sec:Abelian-axial}).
 }
 
 \changed{
In this paper, 
the terminology ``non-Abelian'' is used for vortices 
with non-Abelian magnetic fluxes and 
those accompanied with non-Abelian Nambu-Goldstone 
modes, 
as is common in dense QCD 
\cite{Balachandran:2005ev, Eto:2013hoa}, 
supersymmetric QCD \cite{Hanany:2003hp,Auzzi:2003fs,
Hanany:2004ea,Shifman:2004dr,Eto:2004rz,Eto:2005yh}  
(see Refs.~\cite{Tong:2005un,Eto:2006pg,Shifman:2007ce,Shifman:2009zz} as a review), 
and two-Higgs doublet models 
\cite{Eto:2018hhg,Eto:2018tnk,
Eto:2019hhf,Eto:2020hjb,Eto:2020opf}. 
}

\changed{
Global analogues are also called non-Abelian.
In this paper, vortices winding in chiral symmetry breaking 
$U(N)_{\rm L} \times U(N)_{\rm R} \to SU(N)_{\rm L+R}$ 
are called non-Abelian axial vortices,
see Sec.~\ref{sec:NAGV}.
}

%\begin{textcolor}{red}
\changed{
However note that 
the same terminology ``non-Abelian'' is sometimes 
used for a different meaning in the literature.
It is used for vortices 
with non-Abelian holonomies in Refs.~\cite{Alford:1990mk, Alford:1990ur, Alford:1992yx},
which differs from our terminology.
Note that the above mentioned non-Abelian vortices 
in dense QCD, SUSY QCD and two-Higgs doublet models 
are {\it Abelian} 
in this language since holonomies are ${\mathbb Z}_N$ 
($N=3$ for dense QCD and $N=2$ for two-Higgs doublet models). 
%of  Refs.~\cite{Alford:1990mk, Alford:1990ur, Alford:1992yx}. 
Chiral non-Abelian vortices found in this paper are accompanied by 
non-Abelian holonomies, and thus they are non-Abelian 
in this language as well.
%of Refs.~\cite{Alford:1990mk, Alford:1990ur, Alford:1992yx} as well.
%\end{textcolor}{red}
}

\subsection{Superfluid/semi-superfluid}

\changed{
Abelian superfluid vortices have integer windings around
$U(1)_{\rm B}$. 
They do not carry any color magnetic fluxes.
}

\changed{
Semi-superfluid vortices have fractional 
windings around $U(1)_{\rm B}$. For single-valuedness, 
they must be accompanied with 
color gauge transformation
 for single-valuedness of fields, and thus
 they are inevitably non-Abelian.
  }
  
\subsection{Chiral}

\changed{
We call vortices ``chiral'' 
when only $\Phi_{\rm L}$ or $\Phi_{\rm R}$ has windings.
We label it by $(1,0)$ or $(0,1)$. 
}

\subsection{Topological obstruction}
\label{sec:topological-obstruction}
\changed{
Here, we explain the topological obstruction
\cite{Schwarz:1982ec,
  Alford:1990mk, Alford:1990ur, Alford:1992yx, Preskill:1990bm, 
  Bucher:1992bd, Lo:1993hp,  
Bolognesi:2015mpa}. 
When a symmetry $G$ is spontaneously broken 
down to its subgroup $H$, the OPS is a coset space $G/H$.
Note that the unbroken symmetry $H$ is not unique. 
When the VEV $v = \left<\phi\right>$ of a field $\phi$ is transformed to $v' = g v$ with 
some group element $g \in G$, 
the unbroken symmetry $H$ is also transformed 
to $H' = g H g^{-1}$. 
}

\changed{
A problem may happen in the presence of a vortex.
When we put a vortex, 
the asymptotic value of the field $\phi$ 
depends on the azimuthal angle $\varphi$ around the vortex: 
$\phi(\varphi) \sim g(\varphi) v$. 
Around the vortex, 
the unbroken symmetry $H$ depends on 
 the azimuthal angle $\varphi$ as 
 $H_{\varphi} = g(\varphi) H_0 g(\varphi)^{-1}$ 
 with  $H_0$ being $H$ at $\varphi=0$.
From the single-valuedness of $\phi$, 
we have 
$\phi(\varphi =2\pi) = \phi(\varphi=0)$. 
However, this does not neccesary imply 
$g(\varphi=2\pi)   = g(\varphi=0)$. 
In general, 
$g(\varphi=2\pi)  \neq g(\varphi=0)$, 
and thus the unbroken symmetry 
is not single-valued: 
$H_{\varphi=2\pi} \neq H_{\varphi=0} = H_0$.
This is  the topological obstruction.
An example can be found in an Alice string \cite{Schwarz:1982ec}.
}

%%%%%%%%%%%%%%
\section{Order parameter manifolds}\label{sec:OPM}

Let us describe the full OPM in this Appendix.
To this end, we neglect explicit breaking terms, 
$\gamma_{1,2,3}=0$, thus axial and chiral symmetries becoming exact.

The symmetry $G$ acts on 
the condensates $\Phi_{\rm L,R}$, 
which are $N$ by $N$ matrices of complex scalar fields, as 
\beq
 && \Phi_{\rm L} \to 
g_{\rm C} \Phi_{\rm L} 
 \hat U_{\rm L}^\dagger , \quad
  \Phi_{\rm R} \to 
  g_{\rm C} \Phi_{\rm R} \hat U_{\rm R}^\dagger \nonumber\\
  && g_{\rm C} \in SU(N)_{\rm C}, \quad 
  \hat U_{\rm L,R} \in U(N)_{\rm L,R}.
\eeq
\begin{table}
\begin{center}
\begin{tabular}{c|ccc|cc|cc}
	& $SU(N)_{\rm C}$ & $SU(N)_{\rm L}$ & $SU(N)_{\rm R}$ & $U(1)_{\rm l}$ & $U(1)_{\rm r}$ & $U(1)_{\rm B}$ & $U(1)_{\rm A}$\\
\hline
$({\mathbb Z}_N)_{\rm C}$ & $\omega^k$ & 1 & 1 & 1 & 1 & 1 & 1 \\
$({\mathbb Z}_N)_{\rm L}$ & 1 & $\omega^k$ & 1 & 1 & 1 & 1 & 1 \\
$({\mathbb Z}_N)_{\rm R}$ & 1 & 1 & $\omega^k$ & 1 & 1 & 1 & 1 \\
$({\mathbb Z}_N)_{\rm l}$ & 1 & 1 & 1 & $\omega^k$ & 1 & $\omega^{-\frac{k}{2}}$ & $\omega^{-\frac{k}{2}}$ \\
$({\mathbb Z}_N)_{\rm r}$ & 1 & 1 & 1 & 1 & $\omega^k$ & $\omega^{-\frac{k}{2}}$ & $\omega^{\frac{k}{2}}$ \\
$({\mathbb Z}_N)_{\rm B}$ & 1 & 1 & 1 & $\omega^{-k}$ & $\omega^{-k}$ & $\omega^{k}$ & 1 \\
$({\mathbb Z}_N)_{\rm A}$ & 1 & 1 & 1 & $\omega^{-k}$ & $\omega^{k}$ & 1 & $\omega^{k}$ \\
\hline
$({\mathbb Z}_N)_{\rm C+l+r}$ & $\omega^k$ & 1 & 1 & $\omega^k$ & $\omega^k$ & $\omega^{-k}$ & 1 \\
$({\mathbb Z}_N)_{\rm L+l}$ & 1 & $\omega^k$ & 1 & $\omega^{-k}$ & 1 & $\omega^{\frac{k}{2}}$ & $\omega^{\frac{k}{2}}$ \\
$({\mathbb Z}_N)_{\rm R+r}$ & 1 & 1 & $\omega^k$ & 1 & $\omega^{-k}$ & $\omega^{\frac{k}{2}}$ & $\omega^{-\frac{k}{2}}$\\
\hline
$({\mathbb Z}_N)_{\rm C+B}$ & $\omega^k$ & 1 & 1 & $\omega^k$ & $\omega^k$ & $\omega^{-k}$ & 1\\
$({\mathbb Z}_N)_{\rm L+R+B}$ & 1 & $\omega^k$ & $\omega^k$ & $\omega^{-k}$ & $\omega^{-k}$ & $\omega^{k}$ & 1 \\
$({\mathbb Z}_N)_{\rm L-R+A}$ & 1 & $\omega^k$ & $\omega^{-k}$ & $\omega^{-k}$ & $\omega^{k}$ & 1 & $\omega^{k}$ \\
\hline
$({\mathbb Z}_N)_{\rm C+L+R}$ & $\omega^k$ & $\omega^k$ & $\omega^k$ & 1 & 1 & 1 & 1 \\
$({\mathbb Z}_N)_{\rm C-(L+R)+B}$ & $\omega^k$ & $\omega^{-k}$ & $\omega^{-k}$ & $\omega^{2k}$ & $\omega^{2k}$ & $\omega^{-2k}$ & 1\\
\hline
$({\mathbb Z}_2)_{\rm A+B}$ & 1 & 1 & 1 & 1 & 1 & $-1$ & $-1$ \\
\end{tabular}
\caption{Summary table of the discrete symmetries. $\omega$ is the $N$-th root of the unity: $\omega = \exp (2\pi i /N)$. 
Note $({\mathbb Z}_N)_{\rm C+l+r} = ({\mathbb Z}_N)_{\rm C+B}$.
}
\label{tab:ZN}
\end{center}
\end{table}
With taking into account discrete groups, 
$G$ can be faithfully written as 
\beq
 G &=& {SU(N)_{\rm C} \times U(N)_{\rm L} \times U(N)_{\rm R}
 \over ({\mathbb Z}_N)_{\rm C+l+r} } \non 
  &=&
{SU(N)_{\rm C} \times 
U(1)_{\rm l} \times U(1)_{\rm r} \times 
SU(N)_{\rm L} \times SU(N)_{\rm R} \over 
 ({\mathbb Z}_N)_{\rm C+l+r} \times
 ({\mathbb Z}_N)_{\rm L+l} \times 
 ({\mathbb Z}_N)_{\rm R+r}    \label{eq:fullG1}
} 
\eeq
with 
\beq
 U(N)_{\rm L} = 
 {U(1)_{\rm l} \times SU(N)_{\rm L} \over ({\mathbb Z}_N)_{\rm L+l} },
 \quad 
  U(N)_{\rm R} = 
 {U(1)_{\rm r} \times SU(N)_{\rm R} \over ({\mathbb Z}_N)_{\rm R+r} }.
\eeq
Here, the discrete groups ${\mathbb Z}_N$ 
are defined in Table \ref{tab:ZN}, 
and the two $U(1)$ groups can be explicitly written as
\beq
U(1)_{\rm l}: (\Phi_{\rm L},\Phi_{\rm R}) \to \left(e^{-i\theta_{\rm l}} \Phi_{\rm L},\Phi_{\rm R}\right),\quad
U(1)_{\rm r}: (\Phi_{\rm L},\Phi_{\rm R}) \to \left(\Phi_{\rm L}, e^{-i\theta_{\rm r}}\Phi_{\rm R}\right).\label{eq:U(1)s}
\eeq

Let us rewrite the two $U(1)$ groups 
in Eq.~(\ref{eq:U(1)s})
by the baryon and axial $U(1)$ groups as
\beq
U(1)_{\rm B}: (\Phi_{\rm L},\Phi_{\rm R}) \to e^{i\theta_{\rm B}}\left( \Phi_{\rm L},\Phi_{\rm R}\right),\quad
U(1)_{\rm A}: (\Phi_{\rm L},\Phi_{\rm R}) \to \left(e^{i\theta_{\rm A}}\Phi_{\rm L}, e^{-i\theta_{\rm A}}\Phi_{\rm R}\right),
\eeq
where the relation is given by
\beq
\theta_{\rm B} = - \frac{\theta_{\rm l} + \theta_{\rm r}}{2},\quad
\theta_{\rm A} = - \frac{\theta_{\rm l} - \theta_{\rm r}}{2}.
\eeq
Note that 
\beq
 U(1)_{\rm l} \times U(1)_{\rm r} 
 = {U(1)_{\rm B} \times U(1)_{\rm A}
  \over ({\mathbb Z}_2)_{\rm A+B} },
\eeq
where 
$({\mathbb Z}_2)_{\rm A+B}$ generated by 
$(-1,-1) \in U(1)_{\rm B} \times U(1)_{\rm A}$ 
is redundant and must be removed.
Then, the symmetry $G$ acting on the condensates as
\beq
 && \Phi_{\rm L} \to 
 e^{i\theta_{\rm B} +i\theta_{\rm A}} g_{\rm C} \Phi_{\rm L} U_{\rm L}^\dagger , \quad
  \Phi_{\rm R} \to 
   e^{i\theta_{\rm B} - i \theta_{\rm A}} 
  g_{\rm C} \Phi_{\rm R} U_{\rm R}^\dagger \nonumber\\
  && g_{\rm C} \in SU(N)_{\rm C}, \quad 
  U_{\rm L,R} \in SU(N)_{\rm L,R}, \quad
   e^{i\theta_{\rm B}}\in U(1)_{\rm B} ,\quad
    e^{i\theta_{\rm A}} \in U(1)_{\rm A} 
\eeq
can be rewritten as
\beq
G 
&=&
{SU(N)_{\rm C} \times 
U(1)_{\rm B} \times U(1)_{\rm A} \times 
SU(N)_{\rm L} \times SU(N)_{\rm R} \over 
({\mathbb Z}_2)_{\rm A + B} \times
({\mathbb Z}_N)_{\rm C+B} \times 
({\mathbb Z}_N)_{\rm L+R+B} \times 
({\mathbb Z}_N)_{\rm L-R+A} 
} \non
&=& 
{SU(N)_{\rm C} \times 
U(1)_{\rm B} \times U(1)_{\rm A} \times 
SU(N)_{\rm L} \times SU(N)_{\rm R} \over 
({\mathbb Z}_2)_{\rm A + B} \times
({\mathbb Z}_N)_{\rm C+L+R} \times 
({\mathbb Z}_N)_{\rm C-(L+R)+B} \times
({\mathbb Z}_N)_{\rm L-R+A}  
} \label{eq:fullG2}
\eeq
with the discrete groups in the denominator, defined in Table \ref{tab:ZN}.
In Eq.~(\ref{eq:fullG2}), the direct 
product of the two groups have been rewritten 
by taking the product of the former groups as
$({\mathbb Z}_N)_{\rm C+B} \times 
({\mathbb Z}_N)_{\rm L+R+B} =
({\mathbb Z}_N)_{\rm C+L+R} \times 
({\mathbb Z}_N)_{\rm C-(L+R)+B} $ for later convenience.

The unbroken subgroup $H$ on the ground state $\Phi_{\rm L} \sim \Phi_{\rm R} \sim v  {\bf 1}_N$ is
\beq
 H = 
 {SU(N)_{\rm C+L+R} \times ({\mathbb Z}_N)_{\rm C-(L+R)+B} \times 
   ({\mathbb Z}_N)_{\rm L-R+A}  
 \over 
({\mathbb Z}_N)_{\rm C+L+R} \times 
({\mathbb Z}_N)_{\rm C-(L+R)+B} \times
({\mathbb Z}_N)_{\rm L-R+A}  
}
= { SU(N)_{\rm C+L+R} \over ({\mathbb Z}_N)_{\rm C+L+R} },
\eeq
where the same rearrangements of the discrete groups 
with Eq.~(\ref{eq:fullG1}) have been taken in the denominator.

Thus, the full OPM can be obtained as
\beq
 {\cal M} = {G \over H} 
 =   
{SU(N)_{\rm C} \times 
U(1)_{\rm B} \times U(1)_{\rm A} \times 
SU(N)_{\rm L} \times SU(N)_{\rm R} \over 
 SU(N)_{\rm C+L+R} 
\times
({\mathbb Z}_N)_{\rm C-(L+R)+B} \times
({\mathbb Z}_N)_{\rm L-R+A}  \times 
({\mathbb Z}_2)_{\rm A + B} 
}.
\eeq
Note the relation 
\beq
 SU(N)_{\rm L} \times SU(N)_{\rm R} 
 &=&  SU(N)_{\rm L+R} \ltimes  
   { SU(N)_{\rm L} \times SU(N)_{\rm R} \over  SU(N)_{\rm L+R}}\non
  &\simeq& SU(N)_{\rm L+R} \ltimes  SU(N)_{\rm L-R},
\eeq
where $F \ltimes B$ denotes a fiber bundle 
with a fiber $F$ over a base manifold $B$.\footnote{
\changed{
In general, when a Lie group $G$ is spontaneously broken down to 
$H$, the OPS parametrized by Nambu-Goldstone modes 
is a coset space $G/H$. 
In this situation, the original group $G$ can be regarded as 
a (principal) fiber bundle 
$H \ltimes B$
over the base space $B \simeq G/H$ with a fiber $H$. 
In our case, $H = SU(N)_{\rm L+R}$ and 
$G/H = 
 { SU(N)_{\rm L} \times SU(N)_{\rm R} \over  SU(N)_{\rm L+R}}\simeq SU(N)_{\rm L-R}$.
Here, note that $G/H$ is not endowed with a product 
but isomorphic to a Lie group 
that we denote by $SU(N)_{\rm L-R}$.
}
}
We can further rewrite it as
\beq
 {\cal M}
 &=&
 \left[
{U(1)_{\rm B}  \times SU(N)_{\rm C-(L+R)} \over 
({\mathbb Z}_N)_{\rm C-(L+R)+B} }
 \ltimes
{U(1)_{\rm A} \times  SU(N)_{\rm L-R} \over 
({\mathbb Z}_N)_{\rm L-R+A}  } \right] / ({\mathbb Z}_2)_{\rm A + B} 
 \non
 &=& 
 {U(N)_{\rm C-(L+R)+B} \ltimes U(N)_{\rm L-R+A} 
 \over ({\mathbb Z}_2)_{\rm A + B}}
 =
 {{\cal M}_{\rm V} \ltimes {\cal M}_{\rm A}
 \over ({\mathbb Z}_2)_{\rm A + B}}. \label{eq:fullOPM}
\eeq
Here, we have defined  
the OPMs for the vector symmetry breaking 
and for the axial and chiral symmetry breakings by
\beq
&& {\cal M}_{\rm V} 
 \simeq 
 { U(1)_{\rm B} \times SU(N)_{\rm C} \times SU(N)_{\rm L+R} 
 \over 
 ({\mathbb Z}_N)_{\rm C-(L+R)+B}\times SU(N)_{\rm C+L+R} 
 }
 \simeq 
 {U(1)_{\rm B} \times SU(N)_{\rm C-(L+R)} 
 \over ({\mathbb Z}_N)_{\rm C-(L+R)+B}} 
 \simeq U(N)_{\rm C-(L+R)+B} , \non
&& {\cal M}_{\rm A} \simeq 
{ U(1)_{\rm A}  \times SU(N)_{\rm L} \times SU(N)_{\rm R} 
\over 
 ({\mathbb Z}_N)_{\rm L-R+A} \times SU(N)_{\rm L+R} 
}
\simeq 
{U(1)_{\rm A} \times SU(N)_{\rm L-R} 
 \over ({\mathbb Z}_N)_{\rm L-R+A}} 
 \simeq U(N)_{\rm L-R+A} .
\eeq
with coset spaces 
\beq
  SU(N)_{\rm C-(L+R)} \simeq
  {SU(N)_{\rm C} \times SU(N)_{\rm L+R} 
 \over 
SU(N)_{\rm C+L+R} },\quad
SU(N)_{\rm L-R} \simeq
   { SU(N)_{\rm L} \times SU(N)_{\rm R} \over  SU(N)_{\rm L+R}}.
\eeq

%%%%
\section{Similarities and differences  
with vortices in two-flavor dense QCD}\label{sec:two-flavors}

\changed{
Let us make comments 
on possible similarities with 
recently found 
 non-Abelian Alice strings 
 in two-flavor dense QCD.
 }
 
Recently, two-flavor dense QCD relevant for 
quark-hadron continuity was proposed 
\cite{Fujimoto:2019sxg,Fujimoto:2020cho},
consisting of the 2SC condensation of up and down quarks
in addition to a P-wave condensation of down quarks. 
This phase is referred as the 2SC +$\dd$ phase 
and is further classified into deconfined and confined phases of vortices.
In the deconfined phase, 
the most stable vortices are non-Abelian Alice strings 
which are superfluid vortices 
carrying color magnetic fluxes 
\cite{Fujimoto:2020dsa,Fujimoto:2021bes,Fujimoto:2021wsr}.
The amount of these color magnetic fluxes are 
half of those of non-Abelian semi-superfluid 
vortices in the CFL phase. 

These are 
non-Abelian analogue of Alice strings 
~\cite{Schwarz:1982ec,
  Alford:1990mk, Alford:1990ur, Alford:1992yx, Preskill:1990bm, 
  Bucher:1992bd, Lo:1993hp}, 
  and in particular are 
  an $SU(3) \times U(1)$ extension of Alice strings in $SU(2) \times U(1)$ gauge theory
  \cite{Chatterjee:2017jsi,
  Chatterjee:2017hya, Chatterjee:2019zwx,Nitta:2020ggi}. 
  
  One of the characteristic features of non-Abelian Alice stings is 
  that unbroken symmetry generators are not globally defined around the strings, 
  and in general they are multi-valued (topological obstruction).
  Another characteristic feature, which is more important, is
  that particles encircling these strings can detect the colors of the strings from infinite distances by 
  color non-singlet AB phases.
  
In the confined phase, non-Abelian Alice strings are 
confined by the so-called AB defects 
\cite{Chatterjee:2019zwx,Chatterjee:2018znk,Nitta:2020ggi}   
appearing to compensate a discontinuity originated from 
non-trivial AB phases of the 2SC condensation 
\cite{Fujimoto:2021wsr}.
As a result of vortex confinement, 
there can exist only baryonic and mesonic bound states of 
the Alice strings, 
which exhibit color singlet AB phases of particles encircling them;
The baryonic bound state consists of three Alice strings with different (red, blue, green) 
color magnetic fluxes with total color canceled out,
which are connected by a domain wall junction  
resulting in a single Abelian superfluid vortex, 
while the mesonic bound state consists of two Alice strings with the same color magnetic fluxes,  
which are connected by a single domain wall 
resulting in a doubly-wound non-Abelian string.
Although the latter carries a color magnetic flux, 
it can exist because of color-singlet AB phases, that is, 
the color cannot be detected 
from infinite distance by AB phases of particles encircling it. 
The amount of the color magnetic flux that 
the mesonic bound state of the Alice string (or doubly-wound non-Abelian vortex) 
in two-flavor quark matter 
is the same with that of a non-Abelian string in the CFL phase. 

Moreover, 
both of 
a mesonic bound state of the Alice string (or doubly-wound non-Abelian vortex) 
in two-flavor quark matter 
and a non-Abelian string in the CFL phase 
exhibit ${\mathbb Z}_3$ color-singlet AB phases of heavy quarks, 
and ${\mathbb Z}_2$ color-singlet generalized AB phases of light quarks.
%These facts suggest that non-Abelian vortices in the CFL phase may be divided to more fundamental elements.
%
%As mentioned above, there exist several similarities between 
%chiral non-Abelian vortices in the CFL phase 
%and non-Abelian Alice strings in the 2SC + $\dd$ phase, 
%such as the topological obstruction and color non-singlet AB phases.
%
However, a crucial difference between them is that 
 non-Abelian Alice strings are confined by the AB defects 
 {\it spontaneously} appearing in the formation of the 2SC condensate 
 while chiral non-Abelian vortices are confined by 
 chiral domain walls existing due to the {\it explicit breaking} 
 (mass and anomaly terms)
 of the axial and chiral symmetries.
Thus, we can summarize that 
a salient distinction is whether the appearance of 
the domain walls confining the vortices is due to 
spontaneous (the 2SC + $\dd$ phase) 
or explicit (the CFL phase) breaking. 

\end{appendix}

\bibliographystyle{apsrev4-1}

%\bibliography{bib_chiral}
%\end{document}
%merlin.mbs apsrev4-1.bst 2010-07-25 4.21a (PWD, AO, DPC) hacked
%Control: key (0)
%Control: author (72) initials jnrlst
%Control: editor formatted (1) identically to author
%Control: production of article title (-1) disabled
%Control: page (0) single
%Control: year (1) truncated
%Control: production of eprint (0) enabled
%

\end{document}